\documentclass[%
 reprint,
amsmath,amssymb,
aps,
pra,
]{revtex4-2}
\usepackage{physics}
\usepackage{enumitem}
\usepackage{amssymb}
\usepackage{graphicx}
\usepackage{dcolumn}

\newcommand{\cb}{\color{black}}
\newcommand{\cbl}{\color{black}}

\usepackage{tikz}
\usepackage{pgfplots}
\newcommand{\coline}[1]{\raisebox{2pt}{\tikz{\draw[#1,solid,line width=0.9pt](0,0) -- (5mm,0);}}}
\usepackage{bm}
\usepackage{xcolor}
\pgfplotsset{compat=1.18} 
\begin{document}

\preprint{APS/123-QED}

\title{\textbf{Photon-Mediated Hybridization and Dissipative Transport in a Cavity-QED Ring-Acceptor Architecture.} 
}%

\author{Stephon Alexander$^1$}
\author{Roger Andrews$^2$}
\author{Oliver Fox$^3$}
\author{Sarben Sarkar$^4$}
\affiliation{$^1$Brown Theoretical Physics Center and Department of Physics, Brown University, Providence, Rhode Island 02912, USA}
\affiliation{$^2$Department of Physics, The University of the West Indies, St. Augustine, Trinidad and Tobago}%
\affiliation{$^3$Department of Physics and Astronomy, University of Exeter, Exeter EX4 4QL, United Kingdom}
\affiliation{$^4$Department of Physics, King’s College London, Strand, London WC2R 2LS, United Kingdom}

\date{\today}

\begin{abstract}

\cb 
We investigate excitation transfer in an engineered cavity-QED transport architecture consisting of an $N$-site donor ring coupled coherently to a central acceptor and driven by a single quantized photon mode. The system evolves under a Lindblad master equation including spontaneous loss and pure dephasing. In the ordered symmetric limit, the dynamics reduce exactly to a photon–bright mode–acceptor trimer, allowing closed-form analytic expressions for transfer efficiencies and mode-resolved losses. We demonstrate that near-unity efficiency arises from photon-mediated hybridization that generates a dark transport channel in which ring population is strongly suppressed. This cavity-induced mechanism bypasses dissipative dark modes of the ring and is distinct from conventional excitonic transport or environmentally assisted quantum transport (ENAQT). Static disorder in photon–ring coupling activates lossy ring modes through hybridization, while intra-ring coupling primarily shifts spectral crossings and can restore efficiency by separating dissipative channels. The model is interpreted as a tunable quantum-optical transport device. Our analytic reduction provides clear design principles for engineered quantum transport networks operating in cavity-QED platforms.

\end{abstract}

\cbl

\maketitle

\section{Introduction}

Coherent transport in structured open quantum systems is a central theme in quantum optics and quantum engineering. Networks of coupled quantum emitters interacting with quantized radiation fields provide a controllable setting in which coherence, dissipation, and hybridization combine to shape excitation flow. Understanding how these ingredients produce efficient transport in the presence of loss and disorder is essential both for fundamental open-system physics and for the design of engineered quantum devices.

In this work we investigate excitation transfer in an engineered cavity-QED transport architecture consisting of a ring of coherently coupled two-level systems connected to a central acceptor and driven by a single quantized photon mode. The system evolves under a Lindblad master equation incorporating spontaneous decay, pure dephasing, and static disorder. In the ordered symmetric limit the dynamics reduce exactly to a three-mode system involving the photon, the fully symmetric collective ring excitation, and the acceptor. This reduction allows closed-form analytic expressions for transfer efficiencies and provides a transparent interpretation of the full numerical results.

\cbl
 Plenio and Huelgar~\cite{Plenio2008} considered a dissipative quantum network and found that on-site dephasing can enhance excitation transfer, an effect that was also found in some classical systems.  Li \textit{et al.}~\cite{Li2022} investigated a chromophore dimer consisting of a donor and acceptor and found that strong noise could allow ENAQT while shutting off vibrationally assisted energy transfer. Also Rebentrost \textit{et al.}~\cite{rebentrost2009environment} investigated how dephasing could be utilized through environmental-assisted quantum transport (ENAQT). They found that optimal transport can be obtained in a Fenna–Matthews–Olson (FMO) light-harvesting system at room temperature by varying the energy detunings and hopping parameters. Chin \textit{et al.}~\cite{Chin2010} considered an FMO light-harvesting system and found efficiencies above 90$\%$ with dephasing-assisted transport. It was found that noise suppressed ineffective pathways thus allowing for increased transfer through more direct channels to the RC. Xiong \textit{et al.}~\cite{xiong2014dephasing} considered the effects of dissipation and dephasing on source-network-drain models, and by simulating wave-packet dynamics found increased dephasing changed motion from superdiffusive to diffusive. Applying this model to an FMO complex, they found dephasing slightly increased the energy-transfer efficiency. 

The effect of disorder on quantum chains is associated with Anderson localization~\cite{anderson1958absence}, which generally suppresses quantum transport. The general effect of disorder on quantum transport has been studied in many other systems~\cite{Novo2016,vznidarivc2013transport,mohseni2013geometrical,zerah2020effects,maier2019environment}. Novo \textit{et al.}~\cite{Novo2016} investigated quantum transport in a disordered Frenkel-exciton Hamiltonian, and found that in the weak dephasing regime, increasing disorder can on average increase transport efficiency to an optimal value. \v{Z}nidari\v{c} and Horvat~\cite{vznidarivc2013transport} studied the conductivity of an XX spin chain with disorder and dephasing, and determined the optimal dephasing strength and how it scales with differing disorder strengths. Mohseni \textit{et al.}~\cite{mohseni2013geometrical} explored geometrical effects in disordered open quantum systems with applications to artificial light-harvesting structures. They found a saturation of the maximum transfer efficiencies for 7 or 14 chromophores, coinciding with the number in the FMO complex and LH2 monomers. Zerah-Harush and Dubi~\cite{zerah2020effects} studied the effect of disorder and dephasing on ENAQT with a tight-binding Hamiltonian, and found that disorder can assist in quantum transport even when the system has undergone localization due to the disorder. Maier \textit{et al.}~\cite{maier2019environment} experimentally studied ENAQT on a 10-qubit network with static disorder and variable dephasing noise. They found that increasing noise moved the system away from Anderson localization allowing for improved quantum transport.

A key feature of the present formulation is the explicit inclusion of a quantized photonic degree of freedom. Many studies of excitation transport assume an initially prepared exciton or incoherent pumping; by contrast, treating the photon mode dynamically provides a microscopically consistent description of excitation injection and reveals qualitatively new transport behaviour. In particular, the photon couples selectively to the fully symmetric (“bright”) collective mode of the ring, producing a hybridized photon–bright mode–acceptor manifold. This hybridization generates a long-lived transport eigenstate in which population on the dissipative ring is strongly suppressed. As a result, excitation is transferred predominantly through a channel that largely bypasses lossy ring modes. \cbl In summary, the main technical contributions of this paper are as follows:
\begin{enumerate}
\item A single quantized photon-mode model describing the light-matter interactions, which goes beyond conventional models with use of a prepared initial excitation;
\item An explicit Lindblad model for a light-driven ring--center system with both structured coherent couplings and site-resolved dissipation and dephasing;
\item An exact reduction in the ordered, symmetric limit to a three-level system (photon, bright mode, acceptor), yielding closed-form expressions for transfer efficiency as a function of detuning and dephasing;
\item A demonstration of how disorder in photon--ring couplings and site energies activates lossy dark modes, degrades performance, and shifts resonance;
\item A cavity-QED transport mechanism for the near-unity transfer efficiency under resonant conditions, highlighting the role of the photon-bright mode-acceptor channel, and lower efficiencies in non-resonant conditions due to the various loss channels of the dark modes;
\item Identification of regimes where dephasing or increased intra-ring coupling mitigates these effects via mode-mixing or spectral separation.
\end{enumerate}

The model should be understood as a tunable quantum-optical transport device rather than a microscopic description of any specific biological complex. Ring geometries and central acceptor structures arise naturally in several contexts, but here they serve primarily as a minimal architecture in which collective bright and dark modes can be cleanly separated and their roles in transport unambiguously identified. The emphasis throughout is, therefore, on general mechanisms of coherent transport in engineered open quantum systems.

The core ingredients of the model: collective light–matter coupling, tunable dissipation, and controllable network geometry are increasingly accessible in experimental platforms such as superconducting qubit arrays~\cite{Wong2019,Pisani2023,Wallraff2004,Blais2021,Koch2007}, trapped-ion systems~\cite{so2024trapped}, and photonic or waveguide QED lattices~\cite{forcherio2014nanoring,adamo2012electron,Xue2017,Blais2004}. In such systems, engineered dissipation and symmetry-protected subspaces play a central role in state transfer and quantum information routing. The mechanisms identified here therefore provide design principles for programmable quantum transport devices in which excitation flow can be directed and protected by cavity-mediated hybridization.

This paper is organized as follows. In Sec. II we introduce the Hamiltonian and Lindblad description for both ordered and disordered systems and derive analytic expressions for transfer efficiency in the symmetric limit. Sec. III presents numerical and analytic results for transfer efficiency as functions of detuning, disorder, and dephasing, and identifies the photon-mediated dark transport channel responsible for high efficiency. Sec. IV develops reduced exactly solvable models that provide interpretive guidance for the full dynamics. Conclusions and outlook are given in Sec. V.
\cbl

\section{\label{sec:level1}Theory}
\begin{figure*}
  \centering\includegraphics[width=\textwidth]{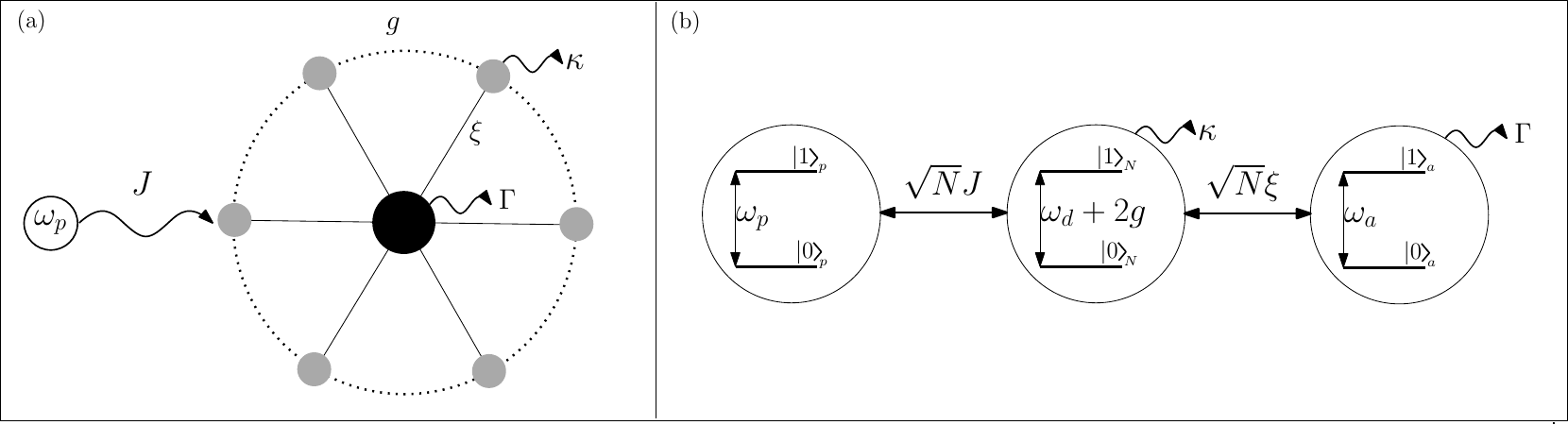}
    \caption{\label{fig:ddpic} Schematic showing the ordered ring-acceptor system with an incident photon. (a) The $N$ donor-ring atoms (gray), with intra-ring coupling $g$, and loss rate $\kappa$ are each coupled to a photon with frequency $\omega_p$ and with coupling constant $J$. Each donor-ring atom is coupled to a central acceptor (black) with coupling constant $\xi$, and loss rate $\Gamma$. (b) A trimer model of the ordered collective ring-acceptor with an incident photon. The photon is coupled to the collective ring with coupling constant $\sqrt{N}J$, and the collective ring is coupled to the acceptor with coupling constant $\sqrt{N}\xi$. Each 2-LS has energy separation $\omega_p$, $\omega_d+2g$, and $\omega_a$, for the photon, collective ring, and acceptor, respectively.}
\end{figure*}

The donor-ring antenna system is made up of $N$ 2-LSs coupled with their nearest-neighbors to create a ring structure (see Fig.~\ref{fig:ddpic}(a)). The Hamiltonian of the ring system, $H_d$ is given as
\begin{equation}
\label{eq:hd}
    H_d=\sum_{j=1}^N\left[\omega_{d,j}e_j^\dagger e_j+g(e^\dagger_je_{j+1}+H.c.)\right],
\end{equation}
where H.c. is Hermitian conjugate, and the annihilation and creation operators, $e_j$ and $e_j^\dagger$, respectively, are defined as
\begin{align}
\label{eq:ejdef}
    e_j=|0_j\rangle\langle1_j|&&e^\dagger_j=|1_j\rangle\langle0_j|,
\end{align}
for $j\in\{1,N\}$, that act on the $j^\text{th}$ 2-LS donor atom. The $j^\text{th}$ 2-LS has states $|0_j\rangle$ and $|1_j\rangle$ with corresponding energies $0$ and $\omega_{d,j}$. The coupling constant between the $j^\text{th}$ and $(j+1)^\text{th}$ 2-LS is $g$.

The acceptor site is a 2-LS with Hamiltonian $H_a$ given as
\begin{eqnarray}
    H_a=\omega_aa^\dagger a,
\end{eqnarray}
where $a^\dagger$ ($a$) is the 2-LS acceptor atom creation (annihilation) operator, and defined as $|1_a\rangle\langle0_a|$ ($|0_a\rangle\langle1_a|$). The acceptor has transition frequency $\omega_a$. The interaction Hamiltonian describing the coupling between the acceptor and the ring, $H_{da}$, is given as
\begin{eqnarray}
\label{eq:hda}
    H_{da}=\xi\sum_{j=1}^N(e_j^\dagger a+e_ja^\dagger),
\end{eqnarray}
where $\xi$ is the donor-acceptor coupling constant

The incident photon has Hamiltonian, $H_p$,
\begin{equation}
\label{eq:hpho}
    H_p=\omega_{p}c_{p}^\dagger c_{p},
\end{equation}
where the photon frequency is $\omega_{p}$ and $p$ defines the mode. The photon annihilation (creation) operator, $c_{p}$ ($c_{p}^\dagger$), are defined as $|1_{p}\rangle\langle0_{p}|$ ($|0_{p}\rangle\langle1_{p}|$). $|0_{p}\rangle$ and $|1_{p}\rangle$ are the vacuum and one photon states, respectively. The photon-ring interaction Hamiltonian is given as
\begin{align}
\label{eq:hdp}
    H_{pd}=\sum_{j=1}^NJ_j(c_{p}^\dagger e_j+c_{p}e_j^\dagger),
\end{align}
where the photon is coupled to each 2-LS on the ring, with $j^{th}$ site-dependent coupling constant $J_j$.

The full system Hamiltonian, $H$, is given as 
\begin{align}
\label{eq:hful}
    H=&H_d+H_a+H_p+H_{pd}+H_{da}\nonumber\\
    =&\sum_{j=1}^N\left(\omega_{d,j}e_j^\dagger e_j+g(e^\dagger_je_{j+1}+H.c.)\right)+\omega_a a^\dagger a\nonumber\\
    &+\omega_{p}c_{p}^\dagger c_{p}+\sum_{j=1}^N\left(J_j(c_{p}^\dagger e_j+c_{p}e_j^\dagger)+\xi(e_j^\dagger a+e_j a^\dagger)\right).
\end{align}

\subsection{Lindblad Master Equation}
For an arbitrary system with Hamiltonian $H'$ coupled to the environment, the Lindblad master equation which describes the time evolution of the density matrix $\rho'$ is as follows~\cite{breuer,ultracold}
\begin{align}
    \dot{\rho'}=-i[H',\rho']+\sum_{n=1}^{N'}\gamma_n\left[L_n\rho' L_n^\dagger-\frac{1}{2}\{L_n^\dagger L_n,\rho'\}\right],
\end{align}
where we have a set of $N'$ jump operators, $L_n$, (along with their Hermitian conjugate $L_n^\dagger$) and with coupling constants $\gamma_n$. The anti-commutator brackets are defined for any pair of operators $A$ and $B$ as $\{A,B\}=AB+BA$. The first term describes the unitary evolution, while the second term describes the incoherent evolution due to coupling with the environment.

For the ring-acceptor system, we consider a Lindblad master equation for the density matrix, $\rho$, of the form
\begin{align}
\label{eq:lbfull}
    \dot{\rho}=-i[H,\rho]+\mathcal{L}_{D}(\rho)+\mathcal{L}_{A}(\rho)+\mathcal{L}_{P}(\rho),
\end{align}
where $H$ is any system Hamiltonian, $\mathcal{L}_{D}$ describes the dissipation via spontaneous emission from the donors to the environment, $\mathcal{L}_{A}$ describes the charge separation from the acceptor for successful transfer, and $\mathcal{L}_{P}$ is a dephasing term of the ring due to its coupling to the environment. These superoperators are defined as follows:
\begin{subequations}
\begin{align}
\label{eq:leqs}
    \mathcal{L}_{D}(\rho)=&\sum_{j=1}^N\kappa(2e_j\rho e_j^\dagger-\{e_j^\dagger e_j,\rho\}),
\end{align}
\begin{align}
    \mathcal{L}_{A}(\rho)=&\Gamma(2a\rho a^\dagger-\{a^\dagger a,\rho\}),
\end{align}
\begin{align}
    \mathcal{L}_{P}(\rho)=&\sum_{j=1}^N\gamma(2e_j^\dagger e_j\rho e_j^\dagger e_j-\{e_j^\dagger e_j,\rho\}),
\end{align}
\end{subequations}
\textcolor{black}{where $2\kappa$ is the spontaneous decay rate of the population of each donor atom on the ring, $2\Gamma$ is the spontaneous decay rate of the population of the acceptor for successful transfer, and $2\gamma$ is the dephasing rate of each donor atom.}
It should be noted that the model does not include the effect of finite temperature.

\subsection{\label{sec:level2}Collective Description of the Donor Ring}
To solve Eq.~(\ref{eq:lbfull}), we start by considering the case where each donor on the ring is equivalent, which occurs when $\omega_{d,j}=\omega_d$ and $J_j=J$ for every $j\in[1,N]$. We define transition operators $e_j$ and $e_j^\dagger$ in terms of $k-$space collective operators $\Tilde{\textbf{e}}_{k}$ and $\Tilde{\textbf{e}}^\dagger_{k}$ as follows:
\begin{subequations}
\label{eq:kspace}
\begin{equation}
   e_j=\frac{1}{\sqrt{N}}\sum_{k}e^{i2\pi jk/N}\Tilde{\textbf{e}}_{k},
\end{equation}
\begin{equation}
    e_j^\dagger=\frac{1}{\sqrt{N}}\sum_{k}e^{-i2\pi jk/N}\Tilde{\textbf{e}}^\dagger_{k},
\end{equation}
\end{subequations}
where $k=0,...,N-1$. Substituting Eq.~(\ref{eq:kspace}a) and~(\ref{eq:kspace}b) into Eq.~(\ref{eq:hd}) (as well as the implied definition of $e_{j+1}$), we obtain the collective ring Hamiltonian, $\Tilde{H}_d$, as
\begin{equation}
\label{eq:ekenergies}
    \Tilde{H}_{d}=\sum_{k=0}^{N-1}\omega_k\Tilde{\textbf{e}}^\dagger_k\Tilde{\textbf{e}}_k,
\end{equation}
where $\omega_k=\omega_d+2g\cos(k)$ is the energy for each ring mode $k$. It should be noted that only the $k=0$ mode couples to the acceptor and the photon, and $\omega_0=\omega_d+2g$ is the $k=0$ mode collective donor-ring energy. The other modes, $k=1,...,N-1$, are decoupled from the photon and the acceptor. Substituting Eq.~(\ref{eq:kspace}a) and~(\ref{eq:kspace}b) in Eq.~(\ref{eq:hda}) and Eq.~(\ref{eq:hdp}), gives the collective donor-acceptor and photon-donor Hamiltonians, $\Tilde{H}_{da}$  and $\Tilde{H}_{pd}$ respectively as
\begin{subequations}
\begin{align}
\label{eq:hdacoll}
    \Tilde{H}_{da}=\sqrt{N}\xi(\Tilde{\textbf{e}}^\dagger_0 a+\Tilde{\textbf{e}}_0 a^\dagger),
\end{align}
\begin{align}
\label{eq:hpdcoll}
    \Tilde{H}_{pd}=\sqrt{N}J(\Tilde{\textbf{e}}^\dagger_0 c_p+\Tilde{\textbf{e}}_0 c_p^\dagger).
\end{align}
\end{subequations}
The Hamiltonian $H$ in Eq.~(\ref{eq:hful}) can be rewritten using Eqs.~(\ref{eq:ekenergies}),~(\ref{eq:hdacoll}) and~(\ref{eq:hpdcoll}), as the collective Hamiltonian $\Tilde{H}$:
\begin{align}
\label{eq:hamilcol}
    \Tilde{H}=&\sum_{k=0}^{N-1}\omega_k\Tilde{\textbf{e}}^\dagger_k\Tilde{\textbf{e}}_k+\omega_a a^\dagger a+\omega_{p}c_{p}^\dagger c_p\nonumber\\
    &+\sqrt{N}\xi(\Tilde{\textbf{e}}^\dagger_0a+H.c.)+\sqrt{N}J(c_{p}^\dagger\Tilde{\textbf{e}}_0+H.c.).
\end{align}
$\Tilde{H}$ can be written into a coupled Hamiltonian, $\Tilde{H}_c$, and a decoupled Hamiltonian, $\Tilde{H}_D$, as $\Tilde{H}=\Tilde{H}_c+\Tilde{H}_D$, where $\Tilde{H}_c$ and $\Tilde{H}_D$ are defined as
\begin{subequations}
\begin{align}
\label{eq:hc}
    \Tilde{H}_c=&\omega_0\Tilde{\textbf{e}}^\dagger_0\Tilde{\textbf{e}}_0+\omega_a a^\dagger a+\omega_{p}c_{p}^\dagger c_p\nonumber\\
    &+\sqrt{N}\xi(\Tilde{\textbf{e}}^\dagger_0a+H.c.)+\sqrt{N}J(c_{p}^\dagger\Tilde{\textbf{e}}_0+H.c.),
\end{align}
\begin{align}
\label{eq:hu}
    \Tilde{H}_D=\sum_{k=1}^{N-1}\omega_k\Tilde{\textbf{e}}_k^\dagger\Tilde{\textbf{e}}_k.
\end{align}
\end{subequations}
The coupled Hamiltonian, $\Tilde{H}_c$, describes the system shown in the schematic in Fig.~\ref{fig:ddpic}(b), and can be written in matrix form as 
\begin{align}
\label{eq:hmat}
    \Tilde{H}_c=\begin{pmatrix}
        \omega_p&\sqrt{N}J&0\\\sqrt{N}J&\omega_0&\sqrt{N}\xi\\0&\sqrt{N}\xi&\omega_p
    \end{pmatrix},
\end{align}
which acts on the basis states: $\{|1_p,0_N,0_a\rangle$, $|0_p,1_N,0_a\rangle$, $|0_p,0_N,1_a\rangle\}$, in the single-excitation subspace. $|1_p,0_N,0_a\rangle$ is the single photon state, $|0_p,1_N,0_a\rangle$ is the single-excitation donor-ring collective state, and $|0_p,0_N,1_a\rangle$ is the single-excitation acceptor state. We denote the eigenvalues of $\Tilde{H}_c$ as $\epsilon_{i}$ ($i=1,2,3$), and the eigenvalues of $\Tilde{H}_D$ are $\omega_k=\omega_d+2g\cos(2\pi k/N)$ ($k=1,...,N-1$).

For any operator expectation $\langle O\rangle$, its time derivative satisfies
\begin{align}
\label{eq:eomexp}
    \frac{d\langle O(t)\rangle}{dt}=\Tr[O\dot{\rho}],
\end{align}
where $\Tr$ represents the trace operation and $\dot{\rho}$ is defined in Eq.~(\ref{eq:lbfull}) for the Hamiltonian $\Tilde{H}$ in Eq.~(\ref{eq:hamilcol}). Using Eq.~(\ref{eq:eomexp}), we obtain the following coupled equations of motion
for the following operator expectation values: $\langle a^\dagger a\rangle$, $\langle c_p^\dagger c_p\rangle$, $\langle \Tilde{\textbf{e}}_{0}^\dagger \Tilde{\textbf{e}}_{0}\rangle$, $\langle\Tilde{\textbf{e}}_{k}^\dagger\Tilde{\textbf{e}}_{k}\rangle_L$, $\langle \Tilde{\textbf{e}}_{0}^\dagger a\rangle$, $\langle \Tilde{\textbf{e}}_{0}^\dagger c_p\rangle$, and $\langle a^\dagger c_p\rangle$,
\begin{subequations}
\label{eq:kspaceeom}
\begin{align}
    \frac{d\langle a^\dagger a\rangle}{dt}=-\Gamma\langle a^\dagger a\rangle+i\sqrt{N}\xi(\langle \Tilde{\textbf{e}}_{0}^\dagger a\rangle-\langle a^\dagger \Tilde{\textbf{e}}_{0}\rangle),
\end{align}
\begin{align}
    \frac{d\langle c_p^\dagger c_p\rangle}{dt}=i\sqrt{N}J(\langle \Tilde{\textbf{e}}_{0}^\dagger c_{p}\rangle-\langle c_{p}^\dagger \Tilde{\textbf{e}}_{0}\rangle),
\end{align}
\begin{align}
\label{eq:e0popeq}
    \frac{d\langle \Tilde{\textbf{e}}_{0}^\dagger \Tilde{\textbf{e}}_{0}\rangle}{dt}=-\left(\kappa+\gamma\left(1-\frac{1}{N}\right)\right)\langle \Tilde{\textbf{e}}_{0}^\dagger \Tilde{\textbf{e}}_{0}\rangle+\frac{\gamma}{N}\langle\Tilde{\textbf{e}}_{k}^\dagger\Tilde{\textbf{e}}_{k}\rangle_L\nonumber\\
    +i\sqrt{N}\xi(\langle a^\dagger \Tilde{\textbf{e}}_{0}\rangle-\langle \Tilde{\textbf{e}}_{0}^\dagger a\rangle) +i\sqrt{N}J(\langle c_p^\dagger \Tilde{\textbf{e}}_{0}\rangle-\langle \Tilde{\textbf{e}}_{0}^\dagger c_p\rangle),
\end{align}
\begin{align}
\label{eq:ekpopeq}
    \frac{d\langle\Tilde{\textbf{e}}_{k}^\dagger\Tilde{\textbf{e}}_{k}\rangle_L}{dt}=\left(-\kappa-\frac{\gamma}{N}\right)\langle\Tilde{\textbf{e}}_{k}^\dagger\Tilde{\textbf{e}}_{k}\rangle_L+\gamma\left(1-\frac{1}{N}\right)\langle \Tilde{\textbf{e}}_{0}^\dagger \Tilde{\textbf{e}}_{0}\rangle,
\end{align}
\begin{align}
    \frac{d\langle \Tilde{\textbf{e}}_{0}^\dagger a\rangle}{dt}=\left(-i(\Delta-2g)-\frac{1}{2}(\gamma+\Gamma+\kappa)\right)\langle \Tilde{\textbf{e}}_{0}^\dagger a\rangle\nonumber\\
    +i\sqrt{N}\xi(\langle a^\dagger a\rangle-\langle \Tilde{\textbf{e}}_{0}^\dagger \Tilde{\textbf{e}}_{0}\rangle)+i\sqrt{N}J\langle c_p^\dagger a\rangle,
\end{align}
\begin{align}
    \frac{d\langle \Tilde{\textbf{e}}_{0}^\dagger c_p\rangle}{dt}=\left(-i(\delta-2g)-\frac{1}{2}(\gamma+\kappa)\right)\langle \Tilde{\textbf{e}}_{0}^\dagger c_p\rangle\nonumber\\
    +i\sqrt{N}J(\langle c_p^\dagger c_p\rangle-\langle \Tilde{\textbf{e}}_{0}^\dagger \Tilde{\textbf{e}}_{0}\rangle)+i\sqrt{N}\xi\langle a^\dagger c_p\rangle,
\end{align}
\begin{align}
    \frac{d\langle a^\dagger c_p\rangle}{dt}=\left(i(\Delta-\delta)-\frac{1}{2}\Gamma\right)\langle a^\dagger c_p\rangle\nonumber\\
    +i\sqrt{N}\xi\langle \Tilde{\textbf{e}}_{0}^\dagger c_p\rangle-i\sqrt{N}J\langle a^\dagger\Tilde{\textbf{e}}_{0}\rangle,
\end{align}
\end{subequations}
where $\langle\Tilde{\textbf{e}}_{k}^\dagger\Tilde{\textbf{e}}_{k}\rangle_L$ is defined as
\begin{align}
    \langle\Tilde{\textbf{e}}_{k}^\dagger\Tilde{\textbf{e}}_{k}\rangle_L=\sum_{k\neq0}\langle\Tilde{\textbf{e}}_{k}^\dagger\Tilde{\textbf{e}}_{k}\rangle.
\end{align}
Additionally, $\delta=\omega_p-\omega_d$ is the photon-ring detuning and $\Delta=\omega_a-\omega_d$ is the acceptor-ring detuning, where $\omega_p$ is the photon frequency, $\omega_a$ is the acceptor frequency, and $\omega_d$ the ring-atom frequency. It should be noted that the $k\neq0$ modes only contribute to the previous set of coupled equations when $\gamma\neq0$.

We are interested in the transfer efficiency, $\eta$, which is the probability that the excitation is successfully transferred from the acceptor, and is defined as
\begin{align}
\label{eq:eta}
    \eta=\int_0^{\infty}2\Gamma\langle a^\dagger a(t)\rangle dt.
\end{align}
$\eta$ can be calculated without solving Eqs.~(\ref{eq:kspaceeom}a)-(\ref{eq:kspaceeom}g) explicitly. This technique is presented in Appendix D, where we apply a Laplace transform technique to obtain a general expression for $\eta$. The analytic form of $\eta$ for the system considered can be found in Eq.~(\ref{eq:etanal}). Another similar technique to obtain analytic expressions for operator expectation values using Lyapunov equations to study relaxation and synchronization in
systems subject to correlated noise can be found in~\cite{10.1063/5.0246275}.

\subsection{\label{sec:level3}Ring-Acceptor with Disorder}
In the more generalized case where parameters $\omega_{d,j}$ and $J_j$ are site-dependent, we can no longer use the collective Hamiltonian $\Tilde{H}$ in Eq.~(\ref{eq:hamilcol}). Instead, we employ the Hamiltonian $H$ in Eq.~(\ref{eq:hful}) and using Eq.~(\ref{eq:lbfull}) we obtain the set of coupled equations for the following operator averages: $\langle c_p^\dagger c_p\rangle$, $\langle a^\dagger a\rangle$, $\langle e_{i}^\dagger e_j\rangle$, $\langle e_j^\dagger a\rangle$, $\langle e_j^\dagger c_p\rangle$, and $\langle a^\dagger c_p\rangle$ ($i,j\in[1,N]$),
\begin{subequations}
\label{eq:fulleom}
\begin{align}
    \frac{d\langle c_p^\dagger c_p\rangle}{dt}=i\sum_{i=1}^NJ_i(\langle e_i^\dagger c_{p}\rangle-\langle c_{p}^\dagger e_i\rangle),
\end{align}
\begin{align}
    \frac{d\langle a^\dagger a\rangle}{dt}=-\Gamma\langle a^\dagger a\rangle+i\sum_{i=1}^N\xi(\langle e_i^\dagger a\rangle-\langle a^\dagger e_i\rangle),
\end{align}
\begin{align}
    \frac{d\langle e_{i}^\dagger e_j\rangle}{dt}&=-\left(\kappa+\gamma(1-\delta_{ij})+i(\delta_i-\delta_j)\right)\langle e_{i}^\dagger e_j\rangle\nonumber\\
    &+iJ_i\langle c_p^\dagger e_j\rangle-iJ_j\langle e_i^\dagger c_p\rangle+i\xi(\langle a^\dagger e_j\rangle-\langle a^\dagger c_p\rangle)\nonumber\\
    &+ig(\langle e_{i+1}^\dagger e_j\rangle+\langle e_{i-1}^\dagger e_j\rangle-\langle e_{i}^\dagger e_{j+1}\rangle+\langle e_{i}^\dagger e_{j-1}\rangle),
\end{align}
\begin{align}
    \frac{d\langle e_j^\dagger a\rangle}{dt}=\left(-i\Delta_j-\frac{1}{2}(\gamma+\Gamma+\kappa)\right)\langle e_j^\dagger a\rangle\nonumber\\
    +i\xi\langle a^\dagger a\rangle+iJ_{j}\langle c_p^\dagger a\rangle-i\xi\sum_{i=0}^N\langle e_j^\dagger e_i\rangle\nonumber\\
    +ig(\langle e_{j-1}^\dagger a\rangle+\langle e_{j+1}^\dagger a\rangle),
\end{align}
\begin{align}
    \frac{d\langle e_j^\dagger c_p\rangle}{dt}=(-i\delta_j-\frac{1}{2}(\gamma+\kappa))\langle e_j^\dagger c_p\rangle\nonumber\\
    +iJ_j\langle c_p^\dagger c_p\rangle+i\xi\langle a^\dagger c_p\rangle-i\sum_{i=0}^NJ_i\langle e_j^\dagger e_i\rangle\nonumber\\
    +ig(\langle e_{j-1}^\dagger c_p\rangle+\langle e_{j+1}^\dagger c_p\rangle),
\end{align}
\begin{align}
    \frac{d\langle a^\dagger c_p\rangle}{dt}=\left(i(\omega_a-\omega_p)-\frac{1}{2}\Gamma\right)\langle \Tilde a^\dagger c_p\rangle\nonumber\\
    +i\sum_{i=1}^N\left(\xi\langle e_j^\dagger c_p\rangle-J_{i}\langle a^\dagger e_j\rangle\right),
\end{align}
\end{subequations}
where $\delta_j=\omega_p-\omega_{d,j}$ is the site dependent photon-donor detuning and $\Delta_j=\omega_a-\omega_{d,j}$ is the site dependent acceptor-donor detuning. We can use the method outlined in the supplementary material
to numerically calculate the transfer efficiency defined in Eq.~(\ref{eq:eta}) using the initial condition $\langle c_p^\dagger c_p(0)\rangle=1$.

Using the $k$-space collective operators in Eq.~(\ref{eq:kspace}), we can calculate the probabilities of the $k-$space population expectations for the system with disorder as
\begin{equation}
\label{eq:kspacedisorder}
    \langle\Tilde{\textbf{e}}_{k}^\dagger\Tilde{\textbf{e}}_{k}\rangle=\frac{1}{N}\sum_{m=1}^{N}\sum_{n=1}^{N}e^{i2\pi k(n-m)/N}\langle e^\dagger_me_n\rangle.
\end{equation}
The loss, $L_k$, for each $k$-mode from the ring via spontaneous decay is defined as
\begin{equation}
\label{eq:losseq}
    L_k=\int_{0}^{\infty}2\kappa\langle\Tilde{\textbf{e}}_{k}^\dagger\Tilde{\textbf{e}}_{k}\rangle dt.
\end{equation}
It should be noted that the probability-conservation identity $\eta+\sum_{k=0}^{N-1}L_k=1$, where the transfer efficiency $\eta$ is defined in Eq.~(\ref{eq:eta}).

Throughout this work the ring–acceptor structure is treated as an engineered quantum-optical transport architecture. While ring geometries arise in biological and condensed-matter contexts, the present model is not intended as a microscopic description of any specific photosynthetic complex. Instead it provides a minimal setting in which photon-mediated collective transport and dissipative hybridization can be analysed analytically and numerically.
\cbl
\section{\label{sec:results}Results}
Unless otherwise stated, the following constant parameter values without disorder are $J=0.1\xi$, $g=0.3\xi$, $\Gamma=0.3\xi$, $\kappa=0.3\xi$, $\omega_a=12\xi$, $\omega_d=11.4\xi$, $\Delta=0.6\xi$, and $\xi=10\text{ps}^{-1}$ as in~\cite{Dong2012,Hu1997}. 

In the case of disorder, the same parameter values are used, i.e. $\Gamma=0.3\xi$, $\kappa=0.3\xi$, $g=0.3\xi$, and $\omega_a=12\xi$. We use $J_i/\xi$ ($\omega_{d,i}/\xi$) as normally distributed random variables with mean $\mu_J$ ($\mu_d$) and standard deviation $\sigma_J$ ($\sigma_d$).

\subsection{\label{sec:methods}Methods}

\textcolor{black}{Our systems contain a set of operators expectation values $\langle O_i(t)\rangle$, $i=1,...,N_0$, with time evolution defined by the equation of motion
\begin{equation}
    \frac{dP(t)}{dt}=MP(t),
\end{equation}
where $P(t)=(\langle O_1(t)\rangle,...,\langle O_{N_0}(t)\rangle)^T$ and $M$ is a time independent matrix of dimension $N_0\times N_0$ with eigenvalues $\epsilon_i$ ($i=1,...,N_0$) where $\text{Re}[\epsilon_i]<0$. For the initial conditions $\langle O_i(0)\rangle=1$ ($i\in\{1,...,N_0\}$) and $\langle O_k(0)\rangle=0$ ($k\neq i$), we found that
\begin{equation}
    \int_0^{\infty}\langle O_j(t)\rangle dt=(-1)^{i+j+1}\frac{\text{Det}[M^{M}_{j,i}]}{\text{Det}[M]},
\end{equation}
where $M^{M}_{j,i}$ is the $(j,i)^{th}$ sub-matrix of $M$. This method is used to calculate the transfer efficiencies $\eta$ defined in Eq.~(\ref{eq:eta}) and the losses $L_k$ defined in Eq.~(\ref{eq:losseq}). $\eta$ and $L_k$ were calculated analytically for the ordered system in Sec.~\ref{sec:level2} where $N_0=10$ and numerically in Sec.~\ref{sec:level3} for the system with disorder where in general for $N$ donors, $N_0=N^2+4N+4$.}

\begin{figure}
  \centering\includegraphics[width=0.48\textwidth]{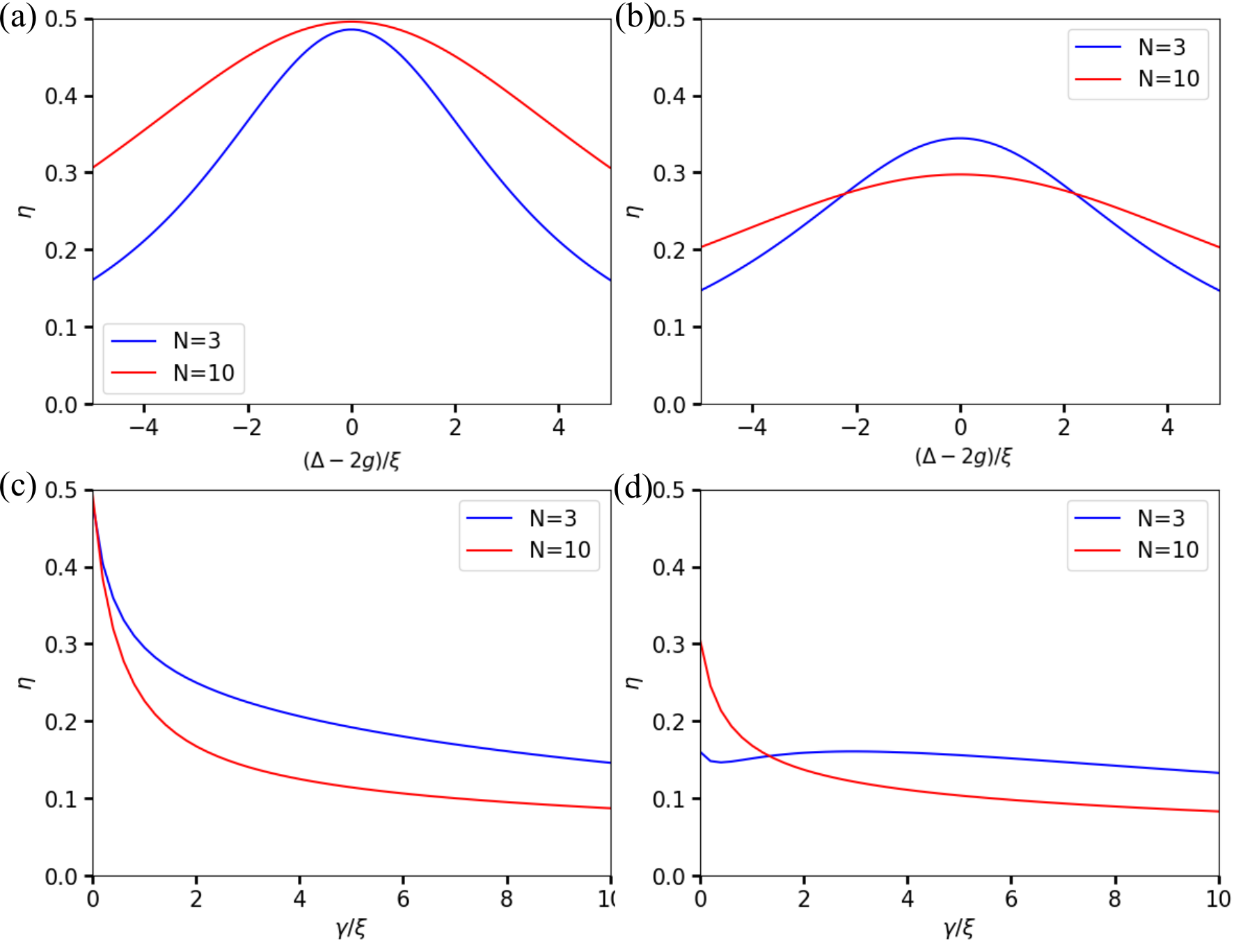}
    \caption{\label{fig:nolight}Plots of the efficiency $\eta_0$ in Eq.~(\ref{eq:etanolig}) for the ring-acceptor system without light and no disorder. (a) Plots of $\eta_0$ against $(\Delta-2g)/\xi$ for $\gamma=0$. (b) Plots of $\eta_0$ against $(\Delta-2g)/\xi$ for $\gamma=0.5\xi$. (c) Plots of $\eta_0$ vs $\gamma/\xi$ with $(\Delta-2g)/\xi=0$. (d) Plots of $\eta_0$ vs $\gamma/\xi$ for $(\Delta-2g)/\xi=5$.}
\end{figure}
\subsection{Ring-Acceptor without Light and no Disorder}

We first consider the case without light and no disorder, 
where the ring has an initial excitation delocalized among the donor atoms. In this case we consider Eqs.~(\ref{eq:kspaceeom}a)-(\ref{eq:kspaceeom}g) with $J=0$ and the initial conditions $\langle \Tilde{\textbf{e}}_{k}^\dagger\Tilde{\textbf{e}}_{k}(0)\rangle=\delta_{0k}$ and $\langle a^\dagger a(0)\rangle=0$. In this case the analytic form of the efficiency, $\eta_0$, calculated using Sec.~\ref{sec:methods}, is as follows
\begin{widetext}
\begin{align}
\label{eq:etanolig}
    \eta_0=\frac{\xi ^2 \Gamma (\gamma +\kappa  N) (\gamma +\kappa +\Gamma)}{\kappa  \Gamma (\gamma +\kappa ) \left((\Delta-2g)^2+(\gamma +\kappa +\Gamma)^2\right)+\xi ^2 (\gamma +\kappa
   +\Gamma) (\kappa  N (\gamma +\kappa +\Gamma)+\gamma  \Gamma)}.
\end{align}
\end{widetext}
It should be noted that for $\gamma/\xi=0$,
\begin{align}
\label{eq:nodeph}
    \eta_0=\frac{\Gamma N \xi^2 (\Gamma+\kappa)}{(\Delta-2g)^2 \Gamma \kappa+(\Gamma+\kappa)^2 \left(\Gamma \kappa+N \xi^2\right)}.
\end{align}
For $(\Delta-2g)/\xi=0$
\begin{align}
    \eta_0=\frac{\xi ^2 \Gamma (\gamma +\kappa  N)}{\xi ^2 (\kappa
    N (\gamma +\kappa +\Gamma)+\gamma 
   \Gamma)+\kappa  \Gamma (\gamma +\kappa )
   (\gamma +\kappa +\Gamma)},
\end{align}
and for $\kappa=0$, $\eta_0=1$

Figs.~\ref{fig:nolight}(a) and~\ref{fig:nolight}(b) are plots of $\eta_0$ against the dimensionless collective acceptor-ring detuning $(\Delta-2g)/\xi$, for dimensionless dephasing values $\gamma/\xi=0$ and $\gamma/\xi=0.5$, respectively for $N=3$ (\coline{blue}) and $N=10$ (\coline{red}). We observe a decrease in the on-resonance maximum transfer efficiency from $0.485$ ($0.496$) to $0.344$ ($0.298$) for $N=3$ ($N=10$) as $\gamma/\xi$ is increased from $\gamma/\xi=0$ to $\gamma/\xi=0.5$.

Figs.~\ref{fig:nolight}(c) and~\ref{fig:nolight}(d) are plots $\eta_0$ against the dimensionless dephasing parameter $\gamma/\xi$, for $(\Delta-2g)/\xi=0$ and $(\Delta-2g)/\xi=5$, respectively.

The key findings are that when there is no dephasing (Eq.~(\ref{eq:nodeph})), larger $N$ always gives larger efficiencies. On-resonant, it is found that increasing the dephasing leads to a monotonic decrease in the transfer efficiency for arbitrary $N$. This decrease is found to be less for the off-resonance case. We should note that our simulations indicate that the decrease is more significant for larger values of $N$. Dephasing couples the $k=0$ mode to the dissipative $k\neq0$ modes, which do not couple to the acceptor.

\begin{figure}
  \centering\includegraphics[width=0.48\textwidth]{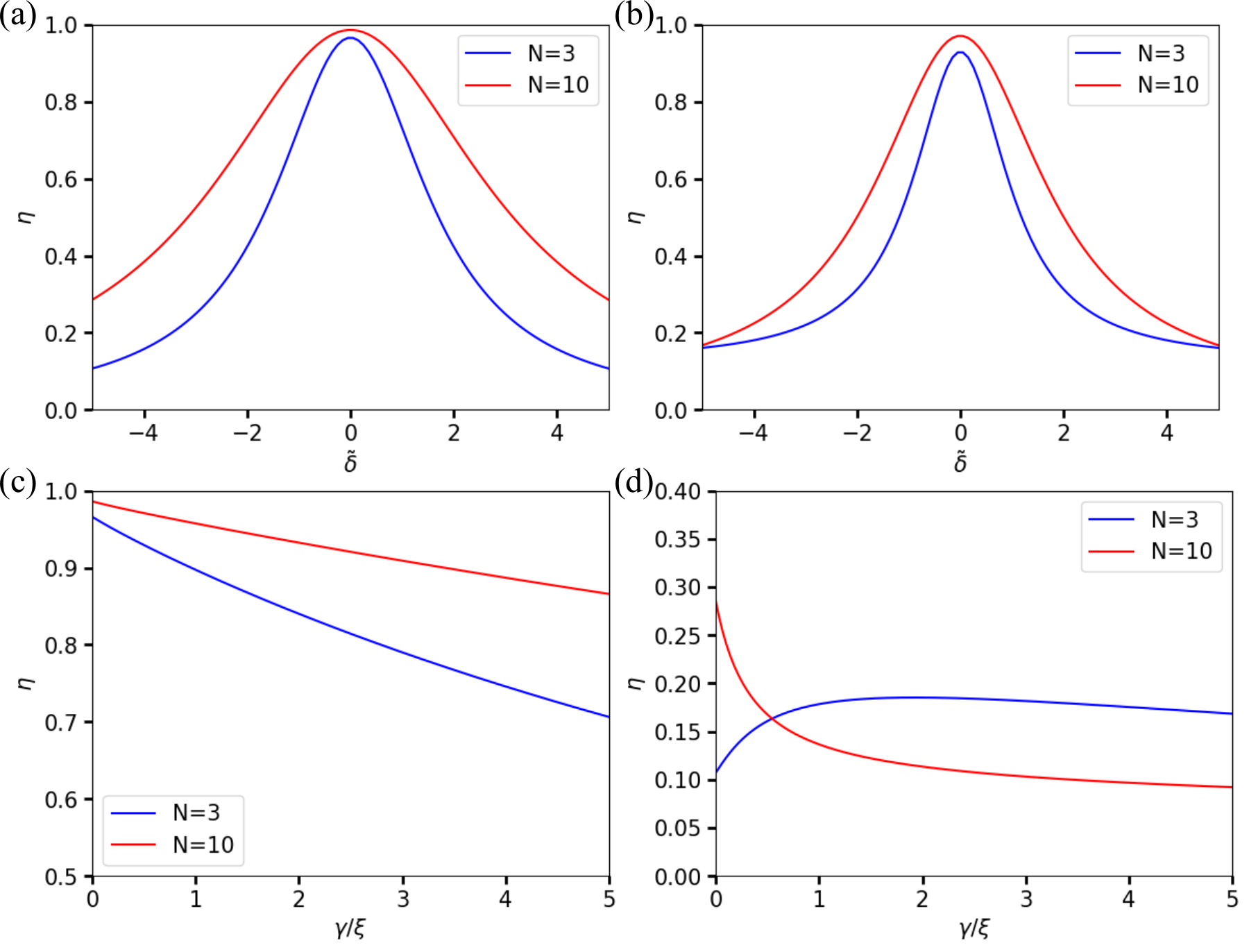}
    \caption{\label{fig:lightanalyt}Plots of the efficiency $\eta$ in Eq.~(\ref{eq:etanal}) with an incident photon and no disorder. (a) Plots of $\eta$ against $\Tilde{\delta}=(\delta-\Delta)/\xi$ for $\gamma/\xi=0$. (b) Plots of $\eta$ against $\Tilde{\delta}/\xi$ for $\gamma/\xi=0.5$. (c) Plots of $\eta$ against $\gamma/\xi$ for $\Tilde{\delta}=0$. (d) Plots of $\eta$ against $\gamma/\xi$ for $\Tilde{\delta}=5$.}
\end{figure}

Dephasing does not act as the primary enhancement mechanism in this architecture. On resonance, increasing dephasing generally reduces efficiency by populating dissipative ring modes. Limited ENAQT-like behaviour appears only in narrow off-resonant regimes where moderate dephasing redistributes population between hybridized eigenstates. The dominant transport mechanism remains photon-mediated hybridization rather than noise-assisted diffusion.
\cbl
\subsection{Ring-Acceptor with an Incident Photon and without Disorder}

Now we consider the case with an incident photon and no disorder. The initial state has $\langle c_p^\dagger c_p(0)\rangle=1$, $\langle \Tilde{\textbf{e}}_{k}^\dagger\Tilde{\textbf{e}}_{k}(0)\rangle=0$, and $\langle a^\dagger a(0)\rangle=0$, and the transfer efficiency is given in Eq.~(\ref{eq:etanal}). Figs.~\ref{fig:lightanalyt}(a) and~\ref{fig:lightanalyt}(b) are plots of $\eta$ against the dimensionless collective photon-ring detuning, $\Tilde{\delta}=(\delta-\Delta)/\xi$, for $\gamma/\xi=0$ and $\gamma/\xi=0.5$, respectively. In both cases ($\gamma/\xi=0$ and $\gamma/\xi=0.5$) we observe a maximum in the transfer efficiency for $N=3$ and $N=10$ at $\Tilde{\delta}/\xi=0$. We observe a small decrease in the on-resonance maximum from $0.966$ ($0.986$) to $0.923$ ($0.971$) for $N=3$ ($N=10$), as dephasing increases from $\gamma/\xi=0$ to $\gamma/\xi=0.5$.

Figs.~\ref{fig:lightanalyt}(c) and~\ref{fig:lightanalyt}(d) are plots of $\eta$ against the dimensionless dephasing parameter $\gamma/\xi$, for $\Tilde{\delta}=0$ and $\Tilde{\delta}=5$, respectively. We observe that for the no detuning case, $\eta$ ($N=10$) $>$ $\eta$ ($N=3$) for all values of $\gamma/\xi$. However, when the detuning is non-zero, $\eta$ ($N=10$) $>$ $\eta$ ($N=3$) occurs in the range $\gamma/\xi<0.546$, and $\eta$ ($N=10$) $<$ $\eta$ ($N=3$) occurs in the range $\gamma/\xi>0.546$.

Comparing the donor-ring acceptor system without light to the system with an incident photon, we observe significantly higher transfer efficiencies in the latter case. For example, for $N=3$, the on-resonance maximum is $0.986$ with an incident photon, and $0.486$ without light. The difference between the transfer efficiencies with and without the photon is due to presence of a dark-state channel with the photon. The dark-state has a very low excitation ring probability and is the dominant state in the excitation transfer process on-resonance. The dark-state is not present without the photon and this leads to a reduction in the transfer efficiency due to enhanced losses from the ring. We also observe that dephasing decreases the maximum transfer efficiency of the system without light more significantly than the system with an incident photon. We note that in Fig. 3(a) when, $N=10$, $\eta>0.9$ for $|\tilde{\delta}|\lessapprox 1$. For $\xi=10 \text{ps}^{-1}$, this gives a bandwidth of the order of $10^{13}$Hz, which far exceeds the bandwidth of a CW laser beam, and therefore the resonance condition is achievable in experiments.

\begin{figure*}[ht]
  \centering\includegraphics[width=0.99\textwidth]{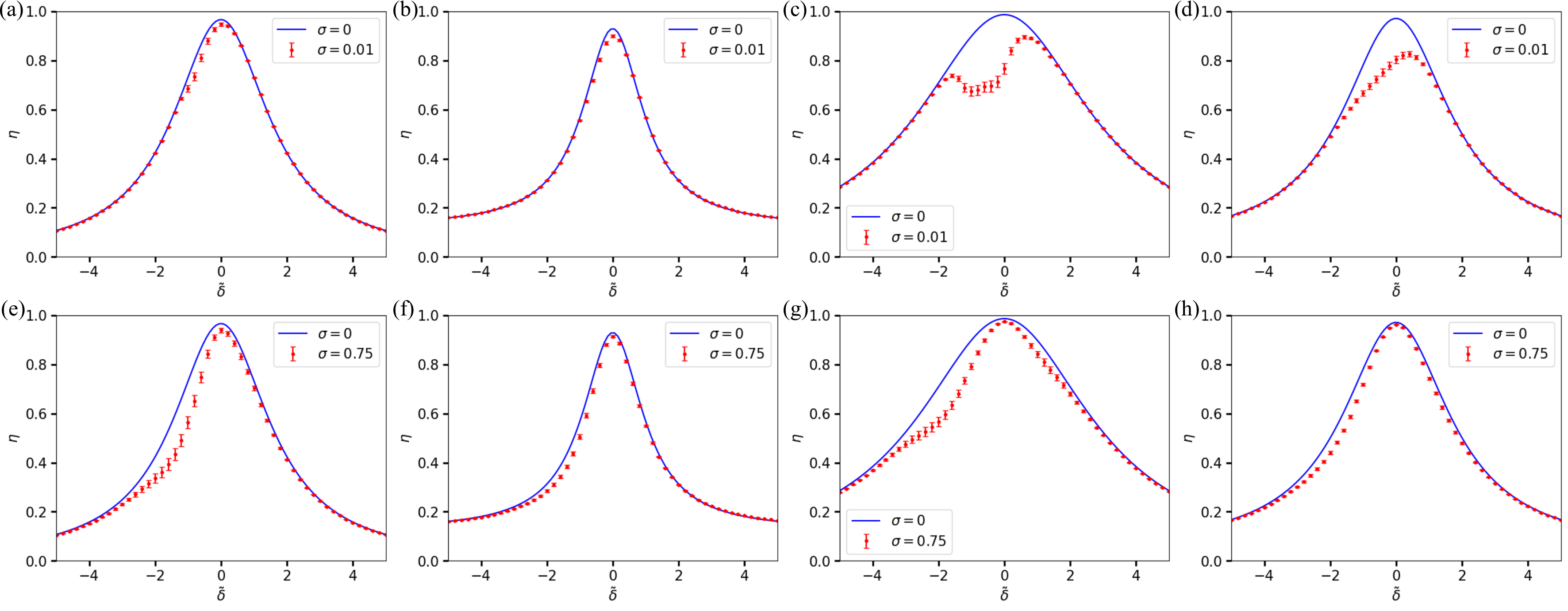}
    \caption{\label{fig:dissj}Plots of the efficiency $\eta$ for the ring-acceptor system with an incident photon and disorder. Figs~\ref{fig:dissj}(a) and~\ref{fig:dissj}(b) (~\ref{fig:dissj}(c) and~\ref{fig:dissj}(d)) are plots of $\eta$ against $\Tilde{\delta}=(\delta-\Delta)/\xi$ for no disorder and disorder in $J_i/\xi$ using $\gamma/\xi=0$ and $\gamma/\xi=0.5$, respectively for $N=3$ ($N=10$). Figs~\ref{fig:dissj}(e) and~\ref{fig:dissj}(f) (\ref{fig:dissj}(g), and~\ref{fig:dissj}(h)) are plots of $\eta$ against $\Tilde{\delta}$ for no disorder and disorder in $\omega_{d,i}/\xi$ for $N=3$ ($N=10$) with $\gamma/\xi=0$ and $\gamma/\xi=0.5$, respectively.}
\end{figure*}
\subsection{Ring-acceptor with an Incident Photon and Disorder in the \textcolor{black}{Energetics} and Photon-Ring Coupling Constants}
\begin{figure*}
  \centering\includegraphics[width=0.99\textwidth]{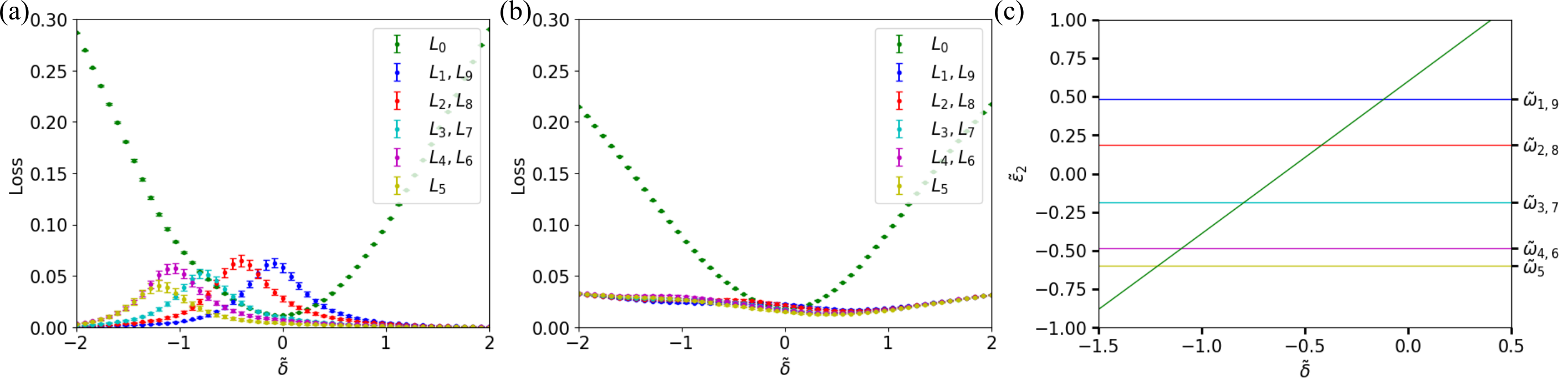}
    \caption{\label{fig:n10eprob}Ring-acceptor system with an incident photon for $g/\xi=0.3$ and $N=10$. (a) and (b) are plots of loss, $L_k$ in Eq.~(\ref{eq:losseq}) against detuning $\Tilde{\delta}=(\delta-\Delta)/\xi$, of collective $k$-modes ($k=0,...,9$) with disorder in $J_i/\xi$, for $\gamma/\xi=0$ and $\gamma/\xi=1$, respectively. (c) Eigenvalue $\Tilde{\epsilon}_2$ of the coupled Hamiltonian $\Tilde{H}_c$ in Eq.~(\ref{eq:hmat}) (\protect\coline{green}\protect) against photon-acceptor detuning $\Tilde{\delta}$. The decoupled energies, $\Tilde{\omega}_k=(\omega_k-\omega_d)/\xi$ ($k=1,...,9$), in Eq.~(\ref{eq:hu}) are marked horizontally.}
\end{figure*}

Figs.~\ref{fig:dissj}(a) and~\ref{fig:dissj}(b) (Figs.~\ref{fig:dissj}(c) and~\ref{fig:dissj}(d)) are plots of $\eta$ against the dimensionless photon-acceptor detuning, $\Tilde{\delta}=(\delta-\Delta)/\xi$, for constant donor-atom onsite energy $\omega_{d,i}=g=11.4\xi$ and disorder in the photon-ring coupling $J_i/\xi$ for $N=3$ ($N=10$), with $\gamma/\xi=0$ and $\gamma/\xi=0.5$, respectively. We use $J_i/\xi$ as a normally distributed random variable with mean $\mu_J=0.1$ and standard deviations $\sigma_J=0$ (\coline{blue}) and $\sigma_J=0.005$ (\coline{red}). $\eta$ is then calculated using the method in Sec.~\ref{sec:methods} for \textcolor{black}{$100$ trials to obtain an average value of $\eta$, with error bars of one standard deviation of the sample using the central limit theorem}. For $N=3$ there is no significant effect of the photon-ring disorder on the transfer efficiency with and without dephasing. For $N=10$, with no dephasing, disorder reduces $\eta$ for $\Tilde{\delta}\in[-2,1.8]$, while for $\gamma/\xi=0.5$, $\eta$ is reduced for $\Tilde{\delta}\in[-1.9,1.4]$. The reduction in $\eta$ is less significant with dephasing. The decrease in efficiency due to disorder is explained by the increased coupling of the photon mode to the $k\neq0$ excitonic modes on the ring, which leads to greater dissipation from the ring, thus lowering the efficiency. Dephasing reduces the negative effect of disorder on the transfer efficiency, because it couples the $k\neq0$ modes to the $k=0$ mode which increases coupling to the acceptor. However, without dephasing the $k\neq0$ modes do not couple to the acceptor and therefore increased dissipation of the excitation from the ring occurs. A more detailed graphical demonstration of dissipation is given in Fig.~\ref{fig:n10eprob}(a) and~\ref{fig:n10eprob}(b).

\textcolor{black}{Figs.~\ref{fig:dissj}(e) and~\ref{fig:dissj}(f) (\ref{fig:dissj}(g) and~\ref{fig:dissj}(h)) are plots of $\eta$ against $\Tilde{\delta}$ for no disorder in $J_i$ ($J_i=0.1\xi$) and disorder in $\omega_{d,i}/\xi$ for $N=3$ ($N=10$), with $\gamma/\xi=0$ and $\gamma/\xi=0.5$, respectively. We use $\omega_{d,i}/\xi$ as a normally distributed random variable with mean $\mu_{\omega_d}=11.4$ and standard deviations $\sigma_{\omega_d}=0$ (\coline{blue}) and $\sigma_{\omega_d}=0.75$ (\coline{red}). Without dephasing for $N=3$ and $N=10$, we find a decrease in transfer efficiency in regions away from resonance, with the decrease being more significant $\tilde{\delta}<0$. With dephasing, we find that disorder in $\omega_{d,i}$ has a negligible effect on the transfer efficiency for all detunings.}

Dips in the transfer efficiency in Figs.~\ref{fig:dissj}(c), \ref{fig:dissj}(d) and \ref{fig:dissj}(g) occur at detuning values when the eigenenergies of the $k\neq0$ ring modes are close to resonance with an eigenenergy of the photon coupled ring-acceptor. These $k\neq0$ modes couple to the photon and this causes additional dissipation from the ring.This effect is demonstrated in more detail in Fig.~\ref{fig:n10eprob} using plots of the loss mode functions $L_k$.

\textcolor{black}{We note that disorder in the parameters $g$, $\gamma$, and $\kappa$ had no effect on the transfer efficiency regardless of other parameter values and disorder strengths. Disorder in $\xi$ caused a decrease in the transfer efficiency equivalent to disorder in $J_i$.}

Figs.~\ref{fig:n10eprob}(a) and~\ref{fig:n10eprob}(b) are plots of the mode losses, $L_k$, defined in Eq.~(\ref{eq:losseq}) for each mode $k=0,...,9$, against the dimensionless detuning $\Tilde{\delta}$ for $N=10$, $\omega_{d,i}/\xi=11,4$ and normally distributed disorder in $J_i/\xi$ ($\mu_J=0.1$ and $\sigma_J=0.01$), for $\gamma/\xi=0$ and $\gamma/\xi=1$, respectively. \textcolor{black}{$\eta$ is calculated for $100$ trials to obtain an average, with error bars of one standard deviation of the sample using the central limit theorem}. Plots are shown for $L_0$ (\coline{green}), $L_1=L_9$ (\coline{blue}), $L_2=L_8$ (\coline{red}), $L_3=L_7$ (\coline{cyan}), $L_4=L_6$ (\coline{purple}) and $L_5$ (\coline{yellow}). 

In Fig.~\ref{fig:n10eprob}(a), $L_0=1-\eta$, where $\eta$ is the efficiency plotted in Fig.~\ref{fig:dissj}(c) in the no-disorder case with no dephasing (\coline{blue}). $L_0$ therefore represents the total loss of the excitation from the ring in the no-disorder and no-dephasing case. $\sum_{k=0}^9L_k=1-\eta$, where $\eta$ is the efficiency plotted in Fig.~\ref{fig:dissj}(c) in the disorder in $J_i/\xi$ case with no dephasing (\coline{red}). Therefore, in the case with disorder, $\sum_{k=0}^9L_k$ represents the total loss of the excitation from the ring. It should be noted that in the case with disorder and no dephasing, the $k$-modes are only coupled to the photon. In addition, the maxima of the various loss terms $L_k$ occur at the following detuning values: $\Tilde{\delta}=-1.2$ for $L_5$; $\Tilde{\delta}=-1.1$ for $L_4$ and $L_6$; $\Tilde{\delta}=-0.8$ for $L_3$ and $L_8$; $\Tilde{\delta}=-0.4$ for $L_2$ and $L_8$; and $\Tilde{\delta}=-0.1$ for $L_1$ and $L_9$.

In Fig.~\ref{fig:n10eprob}(b), with dephasing $\gamma/\xi=1$, the mode losses $L_k$ ($k=1,...,9$) have reduced maxima in the range $-2\leq\Tilde{\delta}\leq2$ compared to the same losses in Fig.~\ref{fig:n10eprob}(a) without dephasing. With disorder and dephasing, each $k$-mode is coupled to the photon and all the other $k$-modes. This leads to a redistribution of losses over the various modes and a flattening of the mode loss terms $L_k$ ($k=1,...,9$). It should be noted that without disorder $L_{k}$ for $k\neq0$ are all equal (see Eq.~(\ref{eq:lkanal}) in Appendix~\ref{app:incoh}).

Fig.~\ref{fig:n10eprob}(c) gives plots of the dimensionless eigenenergy $\Tilde{\epsilon}_{2}=(\epsilon_2-\omega_d)/\xi$ (\coline{green}) against the dimensionless photon-acceptor detuning $\Tilde{\delta}$, for the coupled system without disorder described by the Hamiltonian $\Tilde{H}_c$ in Eq.~(\ref{eq:hc}). The dimensionless eigenenergies, $\Tilde{\omega}_k=(\omega_k-\omega_d)/\xi$ ($k=1,...,9$), of the decoupled system described by the Hamiltonian $\Tilde{H}_D$ in Eq.~(\ref{eq:hu}) are given by the horizontal lines: $\Tilde{\omega}_5=-0.600$ (\coline{yellow}), $\Tilde{\omega}_4=\Tilde{\omega}_6=-0.485$ (\coline{magenta}), 
$\Tilde{\omega}_3=\Tilde{\omega}_7=-0.185$ (\coline{cyan}),
$\Tilde{\omega}_2=\Tilde{\omega}_8=0.185$ (\coline{red}), and
$\Tilde{\omega}_1=\Tilde{\omega}_9=0.485$ (\coline{blue}). The eigenenergy $\Tilde{\epsilon}_2$ coincides with the eigenenergies $\Tilde{\omega}_k$, at the following dimensionless detuning values: $\Tilde{\epsilon}_2=\Tilde{\omega}_5$ at $\Tilde{\delta}=-1.21$; $\Tilde{\epsilon}_2=\Tilde{\omega}_{4,6}$ at $\Tilde{\delta}=-1.10$; 
$\Tilde{\epsilon}_2=\Tilde{\omega}_{3,7}$ at $\Tilde{\delta}=-0.794$; $\Tilde{\epsilon}_2=\Tilde{\omega}_{2,8}$ at $\Tilde{\delta}=-0.419$; and $\Tilde{\epsilon}_2=\Tilde{\omega}_{1,9}$ at $\Tilde{\delta}=-0.116$. These detuning values are approximately equal to the detuning values of the respective peaks in each $L_k$ of Fig.~\ref{fig:n10eprob}(a).

The effect of disorder in $J_i$ introduces the coupling of the decoupled modes of the ring to the photon and these modes decay via spontaneous emission from the ring. The additional coupling of the decoupled modes therefore reduces the transfer efficiency of the system in the detuning range $-2\leq\Tilde{\delta}\leq1.8$ in Fig.~\ref{fig:dissj}(c). This corresponds to the detuning range over which losses occur as shown in Fig.~\ref{fig:n10eprob}(a). The decay of the excitation from the ring via the decoupled modes is greatest when $\Tilde{\omega}_k=\Tilde{\epsilon}_2$, i.e. when the energy of the decoupled mode equals the energy of the first excited state of the collective system. Since $L_0$ is mostly unaffected by disorder, the additional decay channels of the excitation from the donor via the $k\neq0$ modes decreases the transfer efficiency of the excitation.

\textcolor{black}{In the case of disorder in the energies $\omega_{d,i}$, the detuning range in which disorder affects the transfer efficiency suggests a similar mechanism as in the case with disorder in $J_i$. However, instead of additional coupling between the photon and $k\neq0$ modes, we find additional coupling between the $k=0$ and $k\neq0$ modes, when $\tilde{\epsilon}_2\approx\tilde{\omega}_k$. The new channels created by the disorder will only effect the transfer efficiency provided that the donor-ring is sufficiently populated. However, close to resonance, we find a dark-state mechanism whereby the excitation bypasses the ring and thus the disorder in $\omega_{d,i}$ has a negligible effect. We note that the new coupling caused by onsite energetic disorder is similar to that caused by dephasing.}

\begin{figure}
  \centering\includegraphics[width=0.48\textwidth]{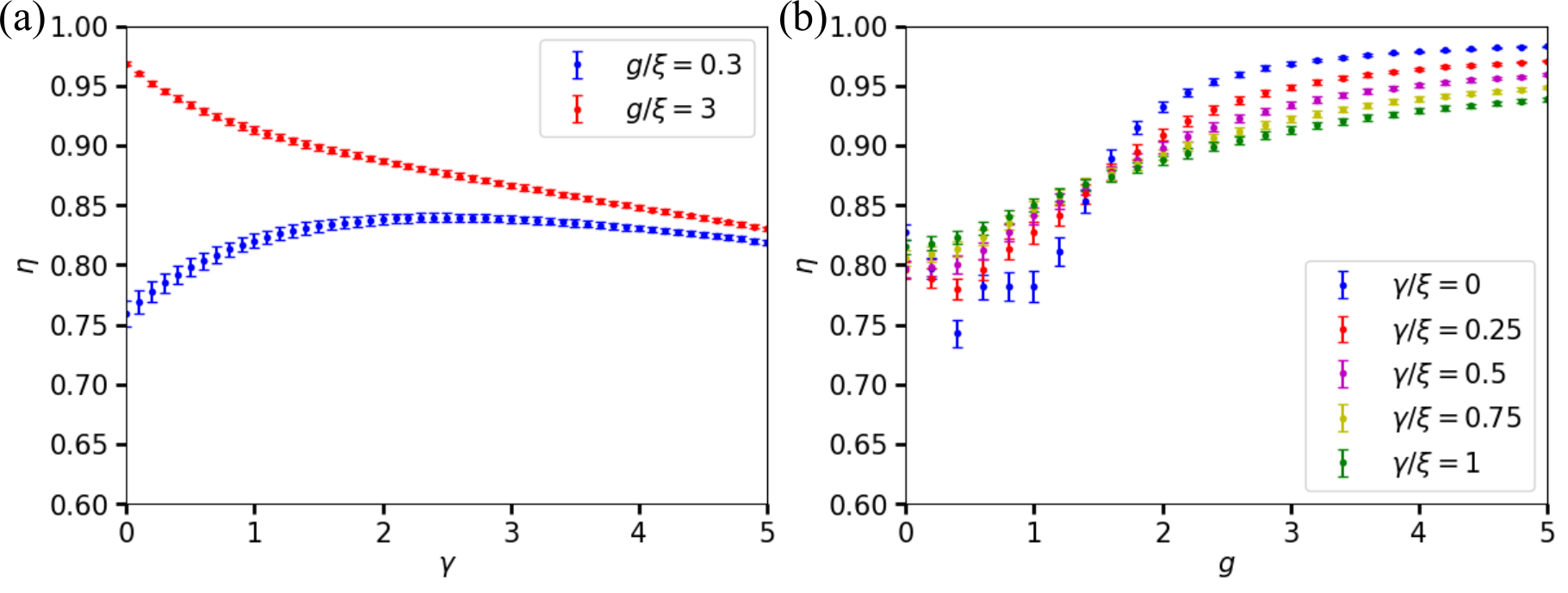}
    \caption{\label{fig:lossdeph}Ring-acceptor system with an incident photon with disorder, for $N=10$, $\Tilde{\delta}=0$, $\omega_{d,i}=\omega_d$ and disorder in $J_i/\xi$. (a) Efficiency $\eta$ against dephasing $\gamma/\xi$. (b) Efficiency $\eta$ against intra-ring coupling $g/\xi$.}
\end{figure}

Fig.~\ref{fig:lossdeph} are plots of the transfer efficiencies for the system with $N=10$, $\Tilde{\delta}=0$, $\omega_{d,i}=\omega_d$, and $J_i/\xi$ normally distributed with $\mu_J=0.1$ and $\sigma_J=0.01$. \textcolor{black}{$\eta$ is calculated for $100$ trials to obtain an average, with error bars of one standard deviation of the sample using the central limit theorem}. Fig.~\ref{fig:lossdeph}(a) are plots of the efficiency $\eta$ against the dimensionless dephasing $\gamma/\xi$ for $g/\xi=0.3$ (\coline{blue}) and $g/\xi=3$ (\coline{red}). We see for $g/\xi=3$, the efficiency has a maximum of $\eta=0.972$ at $\gamma/\xi=0$, and monotonically decreases as $\gamma/\xi$ increases. When $g/\xi=0.3$, as $\gamma/\xi$ increases, $\eta$ initially increases to a maximum of $\eta=0.840$ at $\gamma/\xi=2.5$, followed by a slight monotonic decrease. Fig.~\ref{fig:lossdeph}(b) are plots of the efficiency $\eta$ against intra-ring coupling $g/\xi$ for $\gamma/\xi=0$ (\coline{blue}), $\gamma/\xi=0.25$ (\coline{red}), $\gamma/\xi=0.5$ (\coline{magenta}), $\gamma/\xi=0.75$ (\coline{yellow}) and $\gamma/\xi=1$ (\coline{green}). We see that for $\gamma/\xi=0$, there is an initial decrease to a minimum of $\eta=0.743$ at $g/\xi=0.4$, followed by a monotonic increase. However, for $\gamma/\xi=1$, $\eta$ monotonically increases as $g/\xi$ increases. $\eta$ ($\gamma/\xi=1$) $>$ $\eta$ ($\gamma/\xi=0$) in the range $0\leq g/\xi<1.5$, and $\eta$ ($\gamma/\xi=1$) $<$ $\eta$ ($\gamma/\xi=0$) for $g/\xi>1.5$. In general, when $g/\xi\leq1.5$ an increase in $\gamma/g$ leads to an increase in the transfer efficiency, however when $g/\xi>1.5$ an increase in $\gamma/\xi$ leads to a decrease in the transfer efficiency.

In general, when $g/\xi$ increases, $|\Tilde{\omega}_k|$ increases and $\Tilde{\omega}_k=\Tilde{\epsilon}_i$ occurs at larger values of $|\Tilde{\delta}|$, away from the on-resonance value (see Fig.~\ref{fig:dissj}(c)). The value of $\Tilde{\delta}$ when $\Tilde{\omega}_k=\Tilde{\epsilon}_i$ gives the strongest coupling between the photon and the dissipative $k$-modes of the ring. This means that the photon and the dissipative $k$-modes would have an increasingly weaker coupling at $\Tilde{\delta}=0$ as $g/\xi$ increases, leading to higher transfer efficiencies, which is observed generally in Fig.~\ref{fig:lossdeph}(b). In the no-dephasing case, the $k\neq0$ modes are purely dissipative, so when $g/\xi$ increases, this leads to higher efficiencies at $\Tilde{\delta}=0$. In the case with dephasing, the $k\neq0$ modes couple to the $k=0$ mode, which is coupled to the acceptor. When $g/\xi$ is small, the coupling between the photon and the $k\neq0$ modes is large and dephasing will increase transfer of the excitation to the acceptor. Therefore increasing the dephasing strengthens the coupling between the $k\neq0$ modes and $k=0$ mode thus lowering the population on the dissipative modes, leading to an increase in the transfer efficiency. When $g/\xi$ is large, the coupling between the photon and the $k\neq0$ modes is small and dephasing reduces the transfer efficiency. Increasing dephasing strengthens the coupling between the $k=0$ mode to the dissipative $k$-modes, which decreases the transfer efficiency. We see this in Fig.~\ref{fig:lossdeph}(b), where $g/\xi<1.5$, dephasing has a positive effect on the transfer efficiency, but when $g/\xi>1.5$, it has a negative effect on the transfer efficiency.

\textcolor{black}{We note for general $N$, with $J/\xi\leq0.1$, $\epsilon_2\approx\omega_d+\delta$. Since $\omega_k=\omega_d+2g\cos({2\pi k/N})$, disorder will the decrease the transfer efficiency if $\delta\approx2g\cos({2\pi k/N})$ ($k=1,...,N-1$). In addition, we note that optimal transfer efficiency occurs at $\delta\approx\Delta$. If $|\Delta/\xi|\gtrapprox2$ and we choose $g/\xi\ll1$ then $\delta\gg2g\cos({2\pi k/N})$ and $\eta$ is unaffected by disorder. For $|\Delta/\xi|<2$, $\eta$ will be unaffected by disorder when $\Delta\ll\text{min}[2g\cos({2\pi k/N})]$, which is most optimal when $N=4m+2$ ($m=1,2,...$). In this case, $\text{min}[2g\cos({2\pi k/N})]=2g\sin{\pi/N}$, and we can let $g/\xi\gg\Delta/(2\xi\sin({\pi/N}))$.}

\subsection{The Dark-State Mechanism in the Ring-Acceptor System with an Incident Photon}
\begin{figure}
  \centering\includegraphics[width=0.48\textwidth]{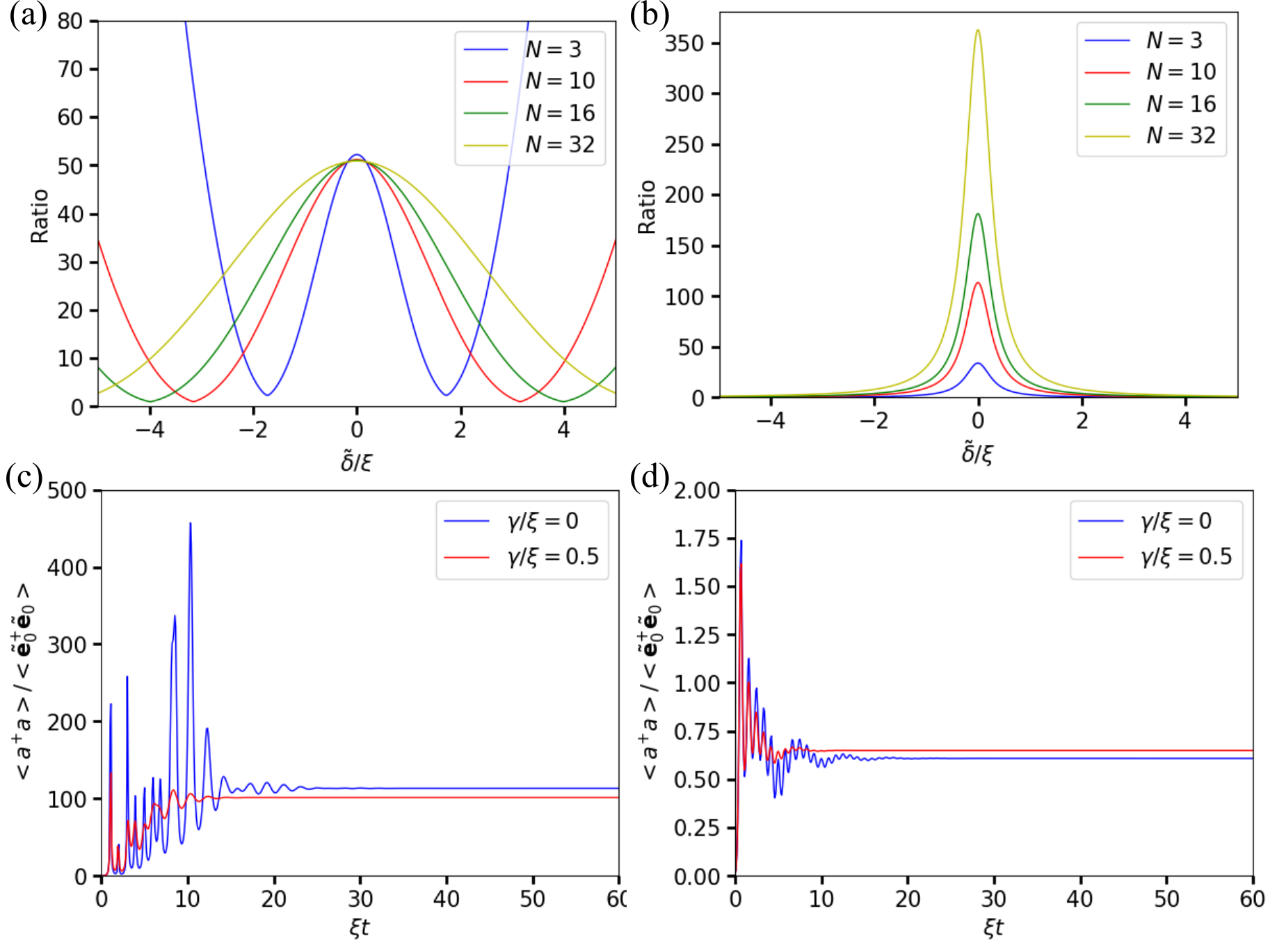}
    \caption{\label{fig:lightprobs}(a) Plot of the ratio of the eigenvalue with the second smallest magnitude for its real part to the eigenvalue with the smallest magnitude for its real part obtained from the matrix in Eq.~(\ref{eq:appenmat}) against $\Tilde{\delta}$. (b) Plot of the population ratio $\langle a^\dagger a\rangle/\langle\Tilde{\textbf{e}}^\dagger_{0}\Tilde{\textbf{e}}_0\rangle$ against $\Tilde{\delta}$ for the eigenvector with eigenvalue whose real part has the smallest magnitude. (c) and (d) are plots of the population ratio $\langle a^\dagger a\rangle/\langle\Tilde{\textbf{e}}^\dagger_{0}\Tilde{\textbf{e}}_0\rangle$ for $N=10$ against $\xi t$ for $\Tilde{\delta}=0$ and $\Tilde{\delta}=4$, respectively.}
\end{figure}
Fig.~\ref{fig:lightprobs} shows plots for the case with no disorder and no dephasing. Fig.~\ref{fig:lightprobs}(a) gives the plot for the ratio of the eigenvalue with the second smallest magnitude of its real part to the eigenvalue with the smallest magnitude of its real part obtained from the matrix in Eq.~\ref{eq:appenmat} against $\Tilde{\delta}$. The set of eigenvalues all have negative real parts which results in exponential decay of the populations on the ring and acceptor. \textcolor{black}{This is done for $N=3$ (\coline{blue}), $N=10$ (\coline{red}), $N=16$ (\coline{green}), and $N=32$ (\coline{yellow}).} An eigenvalue which has a smaller magnitude for its real part would give a longer decay time for the populations. \textcolor{black}{The plots in Fig.~\ref{fig:lightprobs}(a) have maxima of $52.2$, $51,2$, $51,0$, and $50.9$ at $\Tilde{\delta}=0$, for $N=3$, $10$, $16$, and $32$, respectively. When the ratio is large, the eigenvalue whose real part has the smaller magnitude determines the long-time behavior of the populations. When the ratio is small, the long-time behavior is determined by multiple eigenvalues.}

Fig.~\ref{fig:lightprobs}(b) are plots of the population ratio $\langle a^\dagger a\rangle/\langle\Tilde{\textbf{e}}^\dagger_{0}\Tilde{\textbf{e}}_0\rangle$ against $\Tilde{\delta}$ for the eigenvector with eigenvalue whose real part has the smallest magnitude \textcolor{black}{for $N=3$ (\coline{blue}), $N=10$ (\coline{red}), $N=16$ (\coline{green}), and $N=32$ (\coline{yellow}). We see that this ratio has maxima of $34.0$, $113$, $181$, and $362$ at $\Tilde{\delta}=0$ for $N=3$, $10$, $16$, and $32$ respectively.} Therefore these results indicate that when $\Tilde{\delta}$ is small, the long-time dynamics is described by a dark-state, where the excitation probability on the ring is much less than the acceptor. The excitation therefore circumvents the dissipative ring, which reduces loss by spontaneous decay, which results in a high transfer efficiency. \textcolor{black}{For larger $N$, we find that $\langle a^\dagger a\rangle/\langle\Tilde{\textbf{e}}^\dagger_{0}\Tilde{\textbf{e}}_0\rangle\gg 1$ is satisfied for a larger range of detuning values. For example, when $N=32$, we find $\langle a^\dagger a\rangle/\langle\Tilde{\textbf{e}}^\dagger_{0}\Tilde{\textbf{e}}_0\rangle\geq10$ for $|\tilde{\delta}|<1.78$, however when $N=3$, $|\tilde{\delta}|<0.46$. This explains why simulations show that the bandwidth of $\eta$ monotonically increases with $N$ as is also shown in Fig.~\ref{fig:lightanalyt}(a).} It should be noted that Wyke et al.~\cite{wykeroger} found in the perturbative limit ($J/\xi\ll1$) for $\Tilde{\delta}=0$,
\begin{equation}
     \langle a^\dagger a\rangle/\langle\Tilde{\textbf{e}}^\dagger_{0}\Tilde{\textbf{e}}_0\rangle=\frac{N}{(\kappa/\xi)^2},
\end{equation}
which for $N=10$ and $\kappa/\xi=0.3$ gives $\langle a^\dagger a\rangle/\langle\Tilde{\textbf{e}}^\dagger_{0}\Tilde{\textbf{e}}_0\rangle=111$, with the following dark-state
\begin{equation}
    |\psi\rangle\propto e^{-\frac{\sqrt{2} J N \xi}{\kappa^2 + N \xi^2}t}(|P\rangle-\frac{i J \kappa \sqrt{N}}{\kappa^2+N \xi^2}|R\rangle-\frac{J N \xi}{\kappa^2+N
   \xi^2}|A\rangle),
\end{equation}
for $|P\rangle$, $|R\rangle$, and $|A\rangle$ being the excited state of the photon, collective ring and acceptor, respectively. Therefore the dark-state mechanism explains the high transfer efficiency in the system using the phenomenological approach and the Lindblad approach. 

Fig.~\ref{fig:lightprobs}(c) and Fig.~\ref{fig:lightprobs}(d) are plots of the population ratio $\langle a^\dagger a\rangle/\langle\Tilde{\textbf{e}}^\dagger_{0}\Tilde{\textbf{e}}_0\rangle$ for $N=10$ with $\gamma/\xi=0$ (\coline{blue}) and  $\gamma/\xi=0.5$ (\coline{red}) against $\xi t$ for $\Tilde{\delta}=0$ and $\Tilde{\delta}=4$, respectively. In Fig.~\ref{fig:lightprobs}(c), after initial oscillations, $\langle a^\dagger a\rangle/\langle\Tilde{\textbf{e}}^\dagger_{0}\Tilde{\textbf{e}}_0\rangle$ is $113$ and $110$ for $\gamma/\xi=0$ and $\gamma/\xi=0.5$, respectively. The dephasing has a negligible effect on the ratio and hence the dark-state is preserved such that the system maintains a high transfer efficiency. In Fig.~\ref{fig:lightprobs}(d) the long-time population ratios are $0.608$ and $0.649$ for $\gamma/\xi=0$ and $\gamma/\xi=0.5$, respectively. When $|\Tilde{\delta}|\gg0$ the dark-state is destroyed and the excitation no longer circumvents the dissipative ring which leads to a decreased efficiency.

These results do not take into consideration the effect of temperature in the model. Including the effects of temperature in the model would result in damping of the oscillations of the ring and acceptor probabilities, with higher temperatures causing more damping~\cite{PhysRevA.101.062101}. Temperature therefore does not change significantly the decay rates and the dark-state mechanism, so the transfer efficiencies should be qualitatively the same at different temperatures.

The near-unity transfer efficiencies observed in this model arise from photon-mediated hybridization between the photon, the collective bright ring excitation, and the acceptor. This hybridization produces a transport eigenstate with large photon and acceptor amplitudes but strongly suppressed population on dissipative ring modes. Because spontaneous decay acts primarily on the ring degrees of freedom, suppression of ring population dramatically reduces loss and enables highly efficient transfer. The resulting transport channel is therefore a cavity-induced dark pathway rather than a conventional excitonic diffusion or ENAQT mechanism.

The present model employs pure-dephasing Lindblad operators without thermal excitation or detailed-balance transitions. Inclusion of finite-temperature relaxation processes would modify population redistribution among ring modes and is beyond the scope of this work. The results presented here therefore apply most directly to engineered quantum-optical platforms operating effectively at low temperature.
\cbl
\cb
\section{Exactly solvable models as an interpretive scaffold for the numerics}
\label{sec:solvable}

We can give a reduced description of the full ring-acceptor system for which the key dynamical quantities can be obtained in closed form, without large-scale numerics. These models reduce the physics down to the essential couplings while preserving the mechanisms of interest (coupling to light, transfer to the acceptor, and losses). Because they are analytic, they tell us \emph{why} a given numerical trend must occur and give us clean baselines to test our code against.

The donor ring supports collective excitations (normal modes) labelled by a wave number $k$. Only the fully symmetric mode ($k=0$), often called the \emph{bright} mode, couples directly to both the incident photon and the acceptor. All other modes ($k\neq 0$) are \emph{dark}: in the ideal symmetric limit they neither absorb light nor feed the acceptor. This bright/dark decomposition is the central simplification that makes an exact reduction possible.

Projecting the full system onto the bright subspace yields a three-state ``trimer'': (photon) $\leftrightarrow$ (bright ring mode) $\leftrightarrow$ (acceptor). The dark modes are decoupled spectators in the ordered (perfectly symmetric) case. This trimer admits closed-form expressions for the steady transfer efficiency to the acceptor, its line shape as a function of detuning, and its dependence on dephasing. In practice, this trimer is the analytic backbone we use to interpret the full simulations. See Appendix \ref{sec:V-mapping} for details.

\subsection{What the solvable core predicts (and how it guides us)}
\begin{itemize}
  \item \textbf{Tall resonant peak.} When the photon is tuned to the bright--acceptor resonance, the model predicts a large transfer efficiency. This fixes where the main peak \emph{should} sit in the numerics and how tall it can be.
  \item \textbf{Bandwidth narrowing with dephasing.} Adding moderate dephasing reduces the height and narrows the width of the resonant peak: coherence is partially lost, but the bright pathway still dominates. This is a stringent check on the dephasing implementation.
  \item \textbf{Probability bookkeeping.} In the trimer reduction, efficiency $\eta$ plus all loss channels $L_k$ sum to one, $\eta + \sum_k L_k = 1$. This identity is a powerful unit test: if a solver violates it, something is off in the Liouvillian, the rates, or the integrator.
\end{itemize}

\subsection{Disorder as controlled bright--dark mixing}
Static disorder modifies \textcolor{black}{three} things. (i) \emph{Photon--ring coupling disorder} spoils the perfect symmetry by which the photon drives only the bright mode; it leaks amplitude into dark modes that carry loss, especially at detunings where a dark-mode energy crosses a coupled eigenenergy. In numerics this appears as a drop in efficiency and peaks in mode-resolved losses $L_k(\delta)$. \textcolor{black}{(ii) \emph{Energetic disorder} breaks the symmetry and couples the bright mode to the dark modes causing a leakage in amplitude to the dark modes thereby reducing efficiency. Close to resonance the dark state dominates and creates a system which is robust against energetic disorder}. (iii) \emph{Intra-ring coupling disorder} perturbs the dark spectrum but, in typical ranges, has a smaller effect on efficiency because it does not strongly change how light addresses bright vs.\ dark subspaces. The solvable picture makes both trends transparent. 

\subsection{Relation to Spectroscopic Observables}
Two-dimensional electron spectroscopy (2DES), such as the rephasing signal, gives direct experimentally observable signatures of exciton energies, energy transfer pathways and coherence dynamics between the donor ring and the acceptor. For a symmetric system, the bright-mode couples strongly to the acceptor and the dark-modes are suppressed and we should therefore observe a sharp peak on-resonance. With static disorder or dephasing, the bright and dark modes mix, and measurements can be made of the off-diagonal cross peaks and on-resonant peaks for values of detuning where the collective system and dark state energies coincide. These experimental peaks can be compared with the transfer efficiencies to determine how these two quantities are correlated. By adding gain to the donor-ring, it is possible to explore measurements close to exceptional points which may reveal new insights into non-Hermitian physics in ring-acceptor systems.

\vskip .5cm
Exactly solvable reductions (ESR) are more than pedagogical since they provide target curves (peak position, height, and width; monotonic vs.\ non-monotonic trends with dephasing) that the numerical solver must reproduce in the ordered limit before adding complications.
ESR discriminate which features come from the \emph{bright} highway to the acceptor and which arise when population leaks into \emph{dark} `side streets'. This is how we diagnose dark-state transfer vs.\ ordinary hopping.
   We can understand how efficiency scales with detuning, dephasing, ring size, and couplings. This is useful for predicting   where peaks/dips should move when  a parameter is changed.
   When full simulations deviate from analytic slopes, limits, or conservation, we have a clear starting point to find either the bug or identify a regime change.

\cbl
\section{Conclusion}

This paper considered the effects of dephasing and static disorder on a cavity-QED ring-acceptor system using a Lindblad master equation approach. The understanding of the effects of dephasing and disorder on transfer efficiencies in artificial quantum optical ring-acceptor systems is crucial for the design of highly-efficient scalable solar energy devices.

The central physical result of this paper is that near-unity transfer efficiency can arise from this photon-mediated hybridization mechanism. When the photon and acceptor are tuned into resonance with the collective bright mode, the system supports a transport eigenstate with large photon and acceptor amplitudes but only weak ring population. Because spontaneous decay acts primarily on the ring degrees of freedom, suppressing ring population dramatically reduces dissipation and leads to highly efficient transfer. This cavity-induced transport channel is distinct from conventional excitonic diffusion and does not rely on environmentally assisted quantum transport (ENAQT) as a primary mechanism.

\cbl
We considered transfer efficiencies and a cavity-QED transfer mechanism utilizing a dark-state channel in the ring-acceptor system without light and with an incident single photon. Initially, we modelled the no-disorder system with and without dephasing using the collective states of the donor-ring to analytically derive the transfer efficiencies. Then by treating the photon-donor and intra-donor couplings as normally distributed random variables, we numerically obtained the transfer efficiencies with disorder. 

We found that with no disorder, no dephasing, and on-resonance, transfer efficiencies of approximately $50\%$ were found without light, however with an incident photon the peak efficiency was over $98\%$. High efficiencies greater than $95\%$ can be obtained in a narrow photon-acceptor detuning range of $\pm6.23$THz for $N=10$ donor atoms. It is noteworthy that increasing $N$ increases the bandwidth for high transfer efficiencies. These parameter values should be achievable in an engineered quantum optical device. The effect of dephasing causes a significant decrease of the transfer efficiency to $34\%$ without light, however the transfer efficiency remained high at $97\%$ with an incident photon. High efficiencies occur due to the hybridization of the photon-bright mode-acceptor manifold, which creates a channel in which the ring modes are bypassed.

Disorder in photon–ring coupling breaks the perfect symmetry that isolates the bright mode and induces mixing with dissipative dark modes, thereby opening additional loss channels. Energetic disorder and dephasing similarly generate controlled mixing between bright and dark subspaces. By analysing the resulting hybridization structure, we show how efficiency dips arise when hybridized eigenenergies cross dark-mode energies and how increased intra-ring coupling or moderate dephasing can partially mitigate these effects through spectral separation or mode redistribution. One of the limitations of this model is not including the effects of temperature, however we have described qualitatively such effects based on previous work~\cite{PhysRevA.101.062101}.

The principal focus of this work is the identification of a cavity-mediated transport mechanism in which photon–bright–acceptor hybridization generates a long-lived eigenstate with strongly suppressed population on dissipative ring modes. This mechanism enables near-unity excitation transfer in a tunable parameter regime and provides a minimal analytic model of photon-assisted dissipative transport in collective quantum networks. The results are therefore most naturally interpreted in the context of engineered cavity-QED transport devices rather than biological energy-transfer systems.

\cbl

In conclusion, we have investigated the effects of dephasing and disorder on the large transfer efficiency for the artificial ring-acceptor system. The conditions that lead to large transfer efficiencies occur on-resonance with negligible effect of intra-ring coupling disorder. Photon-ring disorder creates extra dissipative channels in the system due to the coupling of dissipative decoupled ring modes to the photon. However increasing the intra-ring coupling reduces the effect of the dissipative decoupled modes on the transfer efficiency. Dephasing enhances the transfer efficiency far away from resonance, but has a negligible effect on the transfer efficiency close to resonance.

\cb The numerical trends reported here follow naturally from a minimal $V$-system mechanism—two donor superpositions feeding a single acceptor/sink—obtained either by truncating the ring to a representative donor dimer $(A,B)$ plus the acceptor $C$ (with bright/dark superpositions $|\pm\rangle$), or by coarse-graining the full ring into a \emph{bright} collective state $|B_0\rangle$, an effective \emph{dark} state $|D\rangle$, and the acceptor $C$. In both viewpoints, only the bright channel couples directly to the acceptor, while the dark channel is decoupled in the symmetric, noise-free limit; moderate dephasing mixes dark back into bright, and static asymmetries or disorder inject amplitude into dark, opening loss channels. This is exactly the mechanism we observe in the full ring simulations with dissipation and disorder.
In the ordered system a symmetric optical drive excites the bright channel exclusively, yielding a near-unity transfer efficiency on resonance. Adding dephasing reduces coherent selectivity and mixes population into lossy dark pathways, narrowing and slightly lowering the peak. In the $V$-model this corresponds to conduction through $|+\rangle\!\leftrightarrow C$ while $|-\rangle$ traps unless mixed by noise.
Far from resonance, purely coherent dynamics suppress transfer; moderate dephasing breaks destructive interference and restores flow to the acceptor, producing the non-monotonic efficiency versus dephasing (a dip then recovery) that we report.

Drive-side disorder (non-uniform photon–ring coupling) is the $V$-model analog of pumping both bright and dark channels: it mixes the input into dark states that do not feed the acceptor and instead dissipate on the ring, reducing efficiency over a detuning window. Mode-resolved loss spectra peak where the dressed bright–acceptor eigenenergy crosses dark-mode energies, confirming this mechanism. By contrast, randomizing intra-ring couplings perturbs the dark spectrum but leaves bright/dark addressing largely intact in the explored range, so the impact on efficiency is much smaller.
In the $V$-language, $g$ sets the bright–dark energy splitting; increasing $g$ pushes dark levels away from the dressed bright–acceptor energy at the operating detuning, reducing resonant leakage into dark and restoring the on-resonance peak. This explains the observed recovery of performance when $g$ is increased.
In the symmetric, single-excitation, narrowband regime, the ring reduces exactly to a photon–bright–acceptor trimer, from which closed-form efficiency line shapes and a probability-conservation identity $\eta+\sum_k L_k=1$ follow. We use these expressions to benchmark the simulations and to predict the location and height of the resonant peak and its dephasing-induced narrowing. A minimal three-state $V$ toy model additionally reproduces coherent trapping and its relief by moderate dephasing.

The $V$-model provides \emph{quantitative} support in the ordered, single-excitation, narrowband limit (where the trimer reduction is exact), and \emph{qualitative} yet predictive guidance with moderate dephasing and weak–to–moderate disorder. For strong disorder, broadband drives, or multi-excitation dynamics, the $V$-model remains a reliable guide to the \emph{direction} of effects—bright/dark balance, and $g$-based mitigation—while the full ring numerics supply the needed quantitative detail.
Taken together, the $V$-model perspective explains why we observe (i) a near-unity resonant peak that narrows under dephasing, (ii) strong sensitivity to photon–ring coupling disorder, \textcolor{black}{moderate sensitivity to energetic disorder}, but weak sensitivity to intra-ring disorder, and (iii) robust recovery of performance by increasing $g$. These points form the core of our conclusions and are consistent across analytic baselines, loss-channel diagnostics, and full numerical results.

\cbl

\begin{acknowledgments}
OIRF is funded by the EPSRC via the Maths DTP 2021-22 University of Exeter (EP/W523859/1).
\end{acknowledgments}

\appendix
\cb
\section{Operator Algebra}
Each donor pigment is modeled as a 2-LS with ground $|g\rangle_j$ and excited $|e\rangle_j$ states.
Writing Pauli ladder operators
$\sigma_j^+ = |e\rangle_{jj}\!\langle g|$, $\sigma_j^- = |g\rangle_{jj}\!\langle e|$, and $\sigma_j^z = \tfrac{1}{2}(|e\rangle_{jj}\!\langle e|-|g\rangle_{jj}\!\langle g|)$,
the \emph{hard-core boson} language expresses the same algebra in terms of bosons that cannot doubly occupy a site.
The Holstein--Primakoff (HP) map \cite{PhysRev.58.1098} gives an exact operator identity for spin-$S$:
\begin{align}
S_j^+ &= b_j^\dagger \sqrt{2S - b_j^\dagger b_j},\\
S_j^- &= \sqrt{2S - b_j^\dagger b_j}\, b_j,\\
S_j^z &=\; S - b_j^\dagger b_j,
\end{align}
where $[b_i,b_j^\dagger]=\delta_{ij}$ are \emph{canonical} bosons.
Identifying $\sigma_j^\pm = S_j^\pm$ and $\sigma_j^z = S_j^z$ with $S=\tfrac{1}{2}$ (2-LS) yields
\begin{align}
\label{eq:a4}
\sigma_j^+ &= b_j^\dagger \sqrt{1 - n_j},\\
\sigma_j^- &= \sqrt{1 - n_j}\, b_j,\\
\sigma_j^z &= \tfrac{1}{2} - n_j,\\
n_j &\equiv b_j^\dagger b_j,
\label{eq:HP_S_half}
\end{align}
where $n_j\in\{0,1\}$. Equations~\eqref{eq:a4}-\eqref{eq:HP_S_half} \emph{enforce} the on-site exclusion ($n_j\in\{0,1\}$) through the square roots, making the HP map \emph{exact} for a 2-LS.
In this sense, hard-core bosons are just canonical bosons with the nonlinearity $\sqrt{1-n_j}$ that forbids double occupancy.

The projected ladder operators $e_j^\dagger\equiv\sigma_j^+$, $e_j\equiv\sigma_j^-$ satisfy
$ \{e_j,e_j^\dagger\}=1$, $e_j^2=(e_j^\dagger)^2=0$ (Pauli/hard-core at a site),
while operators on different sites commute: $[e_i,e_j]=[e_i,e_j^\dagger]=0$ for $i\neq j$.
This is exactly what the HP representation implements when the square roots are kept.

For low excitation density (few excitations spread over many sites), one often \emph{linearizes} HP by expanding $\sqrt{1-n_j}\approx 1 - \tfrac{1}{2}n_j+\cdots$ and retaining only the leading term:
\begin{equation}
\sigma_j^+\approx b_j^\dagger,\qquad \sigma_j^-\approx b_j,\qquad \sigma_j^z\approx \tfrac{1}{2}-n_j.
\end{equation}
This converts the problem into \emph{free bosons}, which is \emph{exact} whenever the Hilbert space is restricted so that no site is ever doubly occupied (e.g.\ the global single-excitation manifold).
Beyond the single-excitation sector, the linearized form \emph{must} be accompanied by either (i) explicit projectors $(1-n_j)$, (ii) the full square roots, or (iii) an infinite on-site repulsion (hard-core limit of a Bose--Hubbard model) to prevent unphysical double occupancies.

Nearest-neighbor exciton transfer terms $J(\sigma_i^+\sigma_j^-+\mathrm{H.c.})$ map under exact HP to
\begin{equation}
J\Big(b_i^\dagger \sqrt{1-n_i}\,\sqrt{1-n_j}\, b_j + \mathrm{H.c.}\Big).
\end{equation}
At low density (or strictly within the single-excitation manifold), $\sqrt{1-n_i}\sqrt{1-n_j}\to 1$ and one recovers the familiar bosonic hopping $J(b_i^\dagger b_j+\mathrm{H.c.})$ \emph{exactly}.
In multi-excitation sectors this number-dependent factor reduces the effective hopping into an already-occupied site and enforces saturation, which is essential for correctly capturing exciton--exciton blocking on a pigment.

With HP exact, population loss from an excited 2-LS site generated by $\mathcal{D}[\sigma_j^-]$ equals that generated by $\mathcal{D}[\sqrt{1-n_j}\,b_j]$.
In linearized HP ($\sigma_j^-\!\approx b_j$), one must ensure the master equation never populates $n_j\ge 2$.
This is automatic in the global single-excitation manifold and remains accurate at low densities, but for double-excitation simulations we recommend either keeping the $\sqrt{1-n_j}$ factors or adding an explicit hard-core constraint (e.g.\ projectors or a large on-site penalty) so that the Lindblad dynamics respect saturation.

\paragraph{When the HP/hard-core picture is exact vs.\ approximate.}
\begin{itemize}
\item \emph{Exact:} any 2-LS description (one exciton per pigment) provided the square roots are retained; all excitation densities; any Hamiltonian/Lindbladian built from $\sigma_j^\pm,\sigma_j^z$.
\item \emph{Exact by restriction:} global single-excitation manifold (at most one excitation in the entire ring): the linearized HP is indistinguishable from exact dynamics because double occupancy cannot occur.
\item \emph{Approximate:} linearized HP in multi-excitation sectors without constraints; accuracy improves as the density per site $\langle n_j\rangle\!\ll\!1$.
\end{itemize}

\paragraph{Generalizations.}
If a pigment admits up to $2S$ local excitations (e.g.\ a three-level ladder), use HP with $S>1/2$:
$S_j^\pm=b_j^\dagger\sqrt{2S-n_j}$, $S_j^z=S-n_j$.
The hard-core case is $S=\tfrac{1}{2}$.
For strictly one-dimensional rings one may alternatively enforce exclusion via a Jordan--Wigner map to fermions; however, for our purposes HP with hard-core (Pauli) operators is both local and directly aligned with the two-level chromophore model.

\section{Explicit mapping to a {$V$} system}
\label{sec:V-mapping}

This section makes explicit the mapping from the ring-acceptor model to the minimal three-level ``$V$ system'' (two upper states feeding one acceptor/sink) introduced earlier, and shows how it captures coherent trapping and disorder-activated losses. We present two equivalent viewpoints: a concrete \emph{site--dimer truncation} and a \emph{collective (bright/dark) coarse-graining}. Throughout we use the Lindblad-rate convention
\begin{equation}
\mathcal{L}[L]\rho \;=\; \Gamma\big(2L\rho L^\dagger-\{L^\dagger L,\rho\}\big),
\,
\label{eq:lindblad-conv}
\end{equation}
so populations decay as $2\Gamma$, and we also write the shorthand dissipator
\begin{equation}
\mathcal{D}[X]\rho \;\equiv\; 2X\rho X^\dagger-\{X^\dagger X,\rho\}.
\end{equation}

\subsection{Site--dimer truncation $\to$ $V$ system (most concrete)}
Pick two representative donors on the ring (sites \(A\) and \(B\)) and the acceptor \(C\).
Restrict the single-excitation Hilbert space to
\begin{equation}
\big\{|A\rangle,\;|B\rangle,\;|C\rangle\big\}.
\end{equation}
In a rotating frame, the Hamiltonian and dissipators read
\begin{align}
H_V
&= \omega_d\!\left(|A\rangle\!\langle A|+|B\rangle\!\langle B|\right) + \omega_a |C\rangle\!\langle C|
 - J\!\left(|A\rangle\!\langle B|\nonumber \right.\\
 &\left.+|B\rangle\!\langle A|\right)
\hspace{.1cm}
 - \xi\!\left(|A\rangle\!\langle C|+|B\rangle\!\langle C| + \text{H.c.}\right),
\label{eq:V-Ham}
\\[2pt]
\dot\rho\big|_{\rm{loss}}
&= \kappa\sum_{s=A,B}\mathcal{D}[|G\rangle\!\langle s|]\rho
\;+\; \Gamma\,\mathcal{D}[|\rm{sink}\rangle\!\langle C|]\rho,
\label{eq:V-loss}
\\
\dot\rho\big|_{\rm{deph}}
&= \gamma\sum_{s=A,B}\mathcal{D}[|s\rangle\!\langle s|]\rho,
\label{eq:V-deph}
\end{align}
with \(\Delta=\omega_a-\omega_d\) the local donor--acceptor detuning. A narrowband photon can be added  (equal-phase drive on A and B).

\paragraph{Bright/dark basis inside the $V$ system.}
Define
\begin{equation}
|+\rangle=\frac{|A\rangle+|B\rangle}{\sqrt{2}},
\qquad
|-\rangle=\frac{|A\rangle-|B\rangle}{\sqrt{2}}.
\end{equation}
Then \(C\) couples only to \(|+\rangle\) with strength \(\sqrt{2}\,\xi\), while \(|-\rangle\) is \emph{dark} (no matrix element to \(C\)).
The A--B tunneling \(J\) sets the internal splitting \(E_\pm=\omega_d\mp J\).
A symmetric optical drive (equal phase on A and B) pumps \(|+\rangle\) exclusively.
Thus the $V$ system exhibits coherent trapping in \(|-\rangle\) at \(\gamma=0\).

\paragraph{Parameter mapping.}
\(J\) is the donor--donor coupling on that bond (ring \(g\) locally);
\(\xi\) is the donor--acceptor coupling;
\(\kappa,\Gamma,\gamma\) are the donor loss, acceptor extraction, and donor dephasing rates (populations decay as \(2\kappa,2\Gamma\) under \eqref{eq:lindblad-conv}).

\subsection{Collective (bright/dark) coarse-graining $\to$ effective $V$ system}
For the ordered ring, define the \emph{bright} collective state
\begin{equation}
|B_0\rangle=\frac{1}{\sqrt{N}}\sum_{j=1}^{N}|j\rangle,
\qquad
\omega_0=\omega_d+2g,
\end{equation}
and choose any orthonormal \emph{dark} superposition \(|D\rangle\) from the \((N-1)\)-dimensional dark subspace (e.g., the lowest-\(|k|\) eigenmode) with energy \(\omega_D\).
Keep the acceptor \(|C\rangle\).
Retaining these three states yields an effective $V$ geometry:
\begin{align}
H_{\text{eff}}
&= \omega_0 |B_0\rangle\!\langle B_0| + \omega_D |D\rangle\!\langle D| + \omega_a |C\rangle\!\langle C|
 - \xi_{\text{eff}}\!|B_0\rangle\!\langle C|\nonumber
 \\
 &+\rm{H.c.}
 + V_{\text{mix}}\!\left(|B_0\rangle\!\langle D|+\rm{H.c.}\right),
\label{eq:Heff}
\end{align}
with \(\xi_{\text{eff}}=\sqrt{N}\,\xi\).
In the perfectly symmetric, noise-free limit \(V_{\text{mix}}=0\).
Static disorder (e.g., photon--ring coupling disorder \(J_j\)) and pure dephasing generate an effective bright--dark mixing \(V_{\text{mix}}\) (coherent or Liouvillian), opening dissipation channels through dark states.

\paragraph{Interpretation.}
The collective mapping is the right language to explain whole-ring numerics: why photon--ring coupling disorder reduces efficiency (it increases effective mixing into dark modes), and why increasing intra-ring coupling \(g\) moves \(\omega_D\) away from the dressed bright--acceptor energy, mitigating loss near the operating detuning.

\subsection{Drives and conventions (consistent with the main text)}
\begin{itemize}
\item \textbf{Optical drive.} In the site--dimer mapping, a symmetric narrowband drive pumps \(|+\rangle\); in the collective mapping it pumps \(|B_0\rangle\).
\item \textbf{Rates.} With \(\mathcal{L}[L]\) as in \eqref{eq:lindblad-conv}, populations on A,B decay at \(2\kappa\), on C at \(2\Gamma\); coherences acquire the usual half-sum of the connected decay rates.
\item \textbf{Detunings.} Local: \(\delta=\omega_p-\omega_d\), \(\Delta=\omega_a-\omega_d\).
Dimensionless: \(\tilde\delta=(\omega_p-\omega_a)/\xi\).
\end{itemize}

\subsection{When the two mappings make the same predictions}
\begin{itemize}
\item \textbf{Ordered, symmetric, on resonance.} Both reduce to a bright-only pathway; the dark channel is decoupled \(\Rightarrow\) tall, narrow efficiency peak.
\item \textbf{Photon--ring coupling \textcolor{black}{and energetic} disorder.} Both represent it as extra coupling into the dark channel (drive imbalance in the dimer picture, nonzero \(V_{\text{mix}}\) in the collective picture), yielding loss bands where \(\omega_D\) crosses the dressed bright--acceptor eigenenergy.
\item \textbf{Increasing \(g\).} Both shift dark energies away from the operating point, mitigating disorder-activated loss and restoring near-peak performance.
\item \textbf{Moderate dephasing.} Both models should exhibit minimal ENAQT off resonance by mixing the dark channel back into the bright pathway; on resonance, dephasing narrows/lowers the peak.
\end{itemize}

\subsection{Quick recipe for analysis and numerics}
\paragraph{V system from a site pair (A,B,C).}
Set \(J=g\), \(\xi,\kappa,\Gamma,\gamma\) as in the ring.
Drive A and B with equal phase; work in the \(|\pm\rangle,C\) basis.
This \(3\times3\) Liouvillian is \emph{exactly solvable} in the single-excitation sector and provides a clean benchmark for trapping.

\paragraph{V system from collective modes ($B_{0}$, D, C).}
Use $\xi_{\rm{eff}}=\sqrt{N}\,\xi$, $\omega_0=\omega_{d}+2g$, and choose $\omega_{D}$ from the ring spectrum.
Model disorder/dephasing as a small bright--dark mixing $V_{\text{mix}}$ (coherent and/or dissipative).
Use this to interpret full-ring numerics and to formulate design rules (e.g., choose $g$ so $\min_{k} \  \left| \omega_{k} -\epsilon_{2\delta_{op}} \right|$ exceeds the broadened linewidth).

\medskip
Either mapping supplies a one-to-one dictionary from the ring-acceptor parameters to a $V$-system parameters, explains coherent trapping, and disorder-activated losses, and yields compact, exactly solvable baselines that validate and interpret the full simulations.

The V model is useful in separating the bright mode from the collective dark mode in the dynamics. The dark mode is treated collectively and the specific behaviour of each individual dark mode is suppressed in this model. This leads to a loss of information about the contribution of each dark mode and their losses to the system. In addition, only specific non-uniformities in the coupling coefficients can be treated analytically in this model. These issues are further compounded in the multi-photon case where analytic results for simple non-uniformities become difficult to determine.

\cbl
\section{\label{app:incoh}Analytic Expressions}
In matrix form, Eqs.~(\ref{eq:kspaceeom}) can be written as
\begin{align}
    \frac{d\textbf{P}(t)}{dt}=\textbf{M}\textbf{P}(t),
\end{align}
where
\begin{widetext}
\begin{subequations}
\begin{align}
\label{eq:appendpvec}
    \textbf{P}(t)=\begin{pmatrix}\langle c_p^\dagger c_p(t)\rangle&\langle a^\dagger a(t)\rangle&\langle \Tilde{\textbf{e}}_{0}^\dagger \Tilde{\textbf{e}}_{0}(t)\rangle&\langle \Tilde{\textbf{e}}_{0}^\dagger c_p(t)\rangle&\langle c_p^\dagger\Tilde{\textbf{e}}_{0}(t)\rangle&\langle a^\dagger c_p(t)\rangle&\langle c_p^\dagger a(t)\rangle&\langle \Tilde{\textbf{e}}_{0}^\dagger a(t)\rangle&\langle a^\dagger\Tilde{\textbf{e}}_{0}(t)\rangle&\langle\Tilde{\textbf{e}}_{k}^\dagger\Tilde{\textbf{e}}_{k}(t)\rangle_L\end{pmatrix}^T,
\end{align}
\begin{align}
\label{eq:appenmat}
    \textbf{M}=\left(
\begin{smallmatrix}0 & 0 & 0 & i J \sqrt{N} & -i J \sqrt{N} & 0 & 0 & 0 & 0 & 0 \\
 0 & -2 \Gamma & 0 & 0 & 0 & 0 & 0 & -i \sqrt{N} \xi  & i \sqrt{N} \xi  & 0 \\
 0 & 0 & -2\left(\kappa+\gamma\left(1-\frac{1}{N}\right)\right) & -i J \sqrt{N} & i J \sqrt{N} & 0 & 0 & i \sqrt{N} \xi  & -i
   \sqrt{N} \xi  & \frac{2 \gamma }{N} \\
 i J \sqrt{N} & 0 & -i J \sqrt{N} & -i \delta_c -\gamma -\kappa  & 0 & i \sqrt{N} \xi  & 0 & 0 & 0 & 0 \\
 -i J \sqrt{N} & 0 & i J \sqrt{N} & 0 & i \delta_c -\gamma -\kappa  & 0 & -i \sqrt{N} \xi  & 0 & 0 & 0 \\
 0 & 0 & 0 & i \sqrt{N} \xi  & 0 & i (\Delta-\delta )-\Gamma & 0 & -i J \sqrt{N} & 0 & 0 \\
 0 & 0 & 0 & 0 & -i \sqrt{N} \xi  & 0 & i (\delta -\Delta)-\Gamma & 0 & i J \sqrt{N} & 0 \\
 0 & -i \sqrt{N} \xi  & i \sqrt{N} \xi  & 0 & 0 & -i J \sqrt{N} & 0 & i \Delta_c-\gamma -\kappa -\Gamma &
   0 & 0 \\
 0 & i \sqrt{N} \xi  & -i \sqrt{N} \xi  & 0 & 0 & 0 & i J \sqrt{N} & 0 & -i \Delta_c-\gamma -\kappa
   -\Gamma & 0 \\
 0 & 0 & \frac{2 \gamma  (N-1)}{N} & 0 & 0 & 0 & 0 & 0 & 0 & -\frac{2\gamma}{N}-2\kappa  \\
\end{smallmatrix}\right),
\end{align}
\end{subequations}
where $\delta_c=\delta-2g$ and $\Delta_c=\Delta-2g$. 

The matrix $\textbf{M}$ can be used to calculate the transfer efficiency $\eta$ in Eq.~(\ref{eq:eta})
\begin{align}
\label{eq:etanal}
    \eta=\frac{\eta_{n}}{\eta_{d}},
\end{align}
where the numerator $\eta_n$ and denominator $\eta_d$ in Eq.~(\ref{eq:etanal}) are defined as
\begin{subequations}
\begin{align}
    \eta_n=&\xi ^2 \Gamma \left(\delta_c^2 \gamma  \mu  \nu -2 \delta_c
   \Delta_c \gamma  \mu  \nu +\Delta_c^2 \mu  (\gamma  \nu +\kappa  N
   \Gamma)+(\gamma  \Gamma+\kappa  N \nu ) \left(\mu  \left(J^2 N+\nu 
   \Gamma\right)+N \nu  \xi ^2\right)\right),
\end{align}
\begin{align}
   \eta_d=&\delta_c^2 \mu  \left(\kappa  \mu  \Gamma \left(\Delta_c^2+\nu ^2\right)+\nu 
   \xi ^2 (\gamma  \Gamma+\kappa  N \nu )\right)-2 \delta_c \Delta_c \mu 
   \left(\kappa  \mu  \Gamma \left(\Delta_c^2-J^2 N+\nu ^2\right)+\nu  \xi ^2
   (\gamma  \Gamma+\kappa  N \nu )\right)\nonumber\\
   &+\Delta_c^2 \mu  \left(\kappa  \mu  \Gamma \left(\gamma ^2+2
   \kappa  (\gamma +\Gamma)+2 \gamma  \Gamma-2 J^2 N+\kappa ^2+2
   \Gamma^2\right)+\xi ^2 \left(\kappa  N \left(\gamma ^2+2 \kappa  (\gamma
   +\Gamma)+2 \gamma  \Gamma+\kappa ^2+2 \Gamma^2\right)+\gamma  \nu 
   \Gamma\right)\right)\nonumber\\
   &+\Delta_c^4 \kappa  \mu ^2
   \Gamma+\Gamma \left(\mu  \left(J^2 N+\nu 
   \Gamma\right)+N \nu  \xi ^2\right) \left(\xi ^2 (\gamma  \Gamma+\kappa  N
   \nu )+\kappa  \mu  \left(J^2 N+\nu  \Gamma\right)\right),
\end{align}
\end{subequations}
where $\delta_c=\delta-2g$, $\Delta_c=\Delta-2g$, $\mu=\kappa+\gamma$ and $\nu=\kappa+\gamma+\Gamma$.

The matrix $\textbf{M}$ can be used to calculate the losses $L_k$ ($k\neq0$) in Eq.~(\ref{eq:losseq}), which are found as
\begin{align}
\label{eq:lkanal}
    L_k=&\gamma  \kappa  \Big(\Delta_c^4 \mu  \Gamma+\delta_c^2 \left(\Delta_c^2
   \mu  \Gamma+\nu ^2 \left(\mu  \Gamma+N \xi ^2\right)\right)+\Delta_c^2
   \left(\mu  \Gamma \left(\gamma ^2+2 \kappa  (\gamma +\Gamma)+2 \gamma 
   \Gamma-2 J^2 N+\kappa ^2+2 \Gamma^2\right)+N \nu ^2 \xi ^2\right)\nonumber\\
   &-2 \Delta_c
   \delta_c \mu  \Gamma \left(\Delta_c^2-J^2 N+\nu ^2\right)-2
   \Delta_c \delta_c N \nu ^2 \xi ^2+\Gamma \left(J^2 N+\nu 
   \Gamma\right) \left(\mu  \left(J^2 N+\nu  \Gamma\right)+N \nu  \xi
   ^2\right)\Big)/\nonumber\\
   &N \Big(\delta_c ^2 \mu  \big(\kappa  \mu  \Gamma \left(\Delta_c^2+\nu
   ^2\right)+\nu  \xi ^2 (\gamma  \Gamma+\kappa  N \nu )\big)-2 \delta_c 
   \Delta_c \mu  \left(\kappa  \mu  \Gamma \left(\Delta_c^2-J^2 N+\nu
   ^2\right)+\nu  \xi ^2 (\gamma  \Gamma+\kappa  N \nu )\right)+\Delta_c^4 \kappa
    \mu ^2 \Gamma\nonumber\\
    &+\Delta_c^2 \mu  \left(\kappa  \mu  \Gamma \left(\gamma
   ^2+2 \kappa  (\gamma +\Gamma)+2 \gamma  \Gamma-2 J^2 N+\kappa ^2+2
   \Gamma^2\right)+\xi ^2 \left(\kappa  N \left(\gamma ^2+2 \kappa  (\gamma
   +\Gamma)+2 \gamma  \Gamma+\kappa ^2+2 \Gamma^2\right)+\gamma  \nu 
   \Gamma\right)\right)\nonumber\\
   &+\Gamma \left(\mu  \left(J^2 N+\nu 
   \Gamma\right)+N \nu  \xi ^2\right) \left(\xi ^2 (\gamma  \Gamma+\kappa  N
   \nu )+\kappa  \mu  \left(J^2 N+\nu  \Gamma\right)\right)\Big),
\end{align}.
\end{widetext}
\section{Collective mode operators and transfer efficiency derivation}

\subsection{\label{sec:coh}Coherent Evolution of Collective Mode Operators}
We consider a trimer system consisting of a single photon coupled to a donor-ring of $N$ 2-level system (2-LS) atoms, and a central 2-LS acceptor. This has Hamiltonian $H$,
\begin{align}
\label{eq:hfulll}
    H=&\sum_{j=1}^N\left(\omega_{d,j}e_j^\dagger e_j+g(e^\dagger_je_{j+1}+H.c.)\right)+\omega_a a^\dagger a\nonumber\\
    &+\omega_{p}c_{p}^\dagger c_{p}+\sum_{j=1}^N\left(J_j(c_{p}^\dagger e_j+c_{p}e_j^\dagger)+\xi(e_j^\dagger a+e_j a^\dagger)\right),
\end{align}
where we have creation (annihilation) operators $e_j^\dagger$ ($e_j$), $c_p^\dagger$ ($c_p$), and $a^\dagger$ ($a$), for the $j^{th}$ ring atom, the photon, and the acceptor, respectively. The energies of the $j^th$ ring atom, photon, and the acceptor are $\omega_{d,j}$, $\omega_p$, and $\omega_a$, respectively. The photon is coupled to the $j^th$ donor atom with coupling constant $J_j$, each donor atom is coupled to its nearest neighbours with coupling constant $g$ and to the acceptor by $\xi$. 

If we assume symmetry between each donor atom, where $\omega_{d,j}=\omega_d$ and $J_j=J$, we can define transition operators $e_j$ and $e_j^\dagger$ in terms of $k-$space collective operators $\Tilde{\textbf{e}}_{k}$ and $\Tilde{\textbf{e}}^\dagger_{k}$ as follows:
\begin{subequations}
\label{eq:kspace}
\begin{equation}
   e_j=\frac{1}{\sqrt{N}}\sum_{k}e^{i2\pi jk/N}\Tilde{\textbf{e}}_{k},
\end{equation}
\begin{equation}
    e_j^\dagger=\frac{1}{\sqrt{N}}\sum_{k}e^{-i2\pi jk/N}\Tilde{\textbf{e}}^\dagger_{k},
\end{equation}
\end{subequations}
where $k=0,...,N-1$. The Hamiltonian $H$ in Eq.~(\ref{eq:hfulll}) can be rewritten as the collective Hamiltonian $\Tilde{H}$:
\begin{align}
\label{eq:hamilcoll}
    \Tilde{H}=&\sum_{k=0}^{N-1}\omega_k\Tilde{\textbf{e}}^\dagger_k\Tilde{\textbf{e}}_k+\omega_a a^\dagger a+\omega_{p}c_{p}^\dagger c_p\nonumber\\
    &+\sqrt{N}\xi(\Tilde{\textbf{e}}^\dagger_0a+H.c.)+\sqrt{N}J(c_{p}^\dagger\Tilde{\textbf{e}}_0+H.c.).
\end{align}
Notably, only the $k=0$ ring mode couples to the photon and the acceptor.

The closed quantum system (without coupling to the environment), has time evolution governed by the Von-Neumann equation through a Hamiltonian $H$, as follows
\begin{equation}
\label{eq:rhodot}
    \frac{d\rho}{dt}=\dot{\rho}=-i[\Tilde{H},\rho],
\end{equation}
where $\rho$ is the system's density matrix.
For any operator expectation $\langle O\rangle$, its time derivative satisfies
\begin{align}
\label{eq:eomexp}
    \frac{d\langle O(t)\rangle}{dt}=\Tr[O\dot{\rho}],
\end{align}
where $\Tr$ represents the trace operation. As we are interested in the time evolution of the populations, we first consider $\langle a^\dagger a\rangle$, the average number of excitations on the acceptor, which, because the acceptor is a 2-LS, is equivalent to the probability that the acceptor is excited. By substituting $\langle a^\dagger a\rangle$ in Eq.~(\ref{eq:eomexp}), we obtain
\begin{align}
    \frac{d\langle a^\dagger a\rangle}{dt}=-i\Tr[a^\dagger a(\Tilde{H}\rho-\rho \Tilde{H})].
\end{align}
We obtain a set of coupled equations for the time evolution of the population expectation values as follows
\begin{subequations}
\label{eq:nolindblad}
\begin{align}
    \frac{d\langle a^\dagger a\rangle}{dt}=i\sqrt{N}\xi(\langle \Tilde{\textbf{e}}_{0}^\dagger a\rangle-\langle a^\dagger \Tilde{\textbf{e}}_{0}\rangle),
\end{align}
\begin{align}
\label{eq:appcpc}
    \frac{d\langle c_p^\dagger c_p\rangle}{dt}=i\sqrt{N}J(\langle \Tilde{\textbf{e}}_{0}^\dagger c_{p}\rangle-\langle c_{p}^\dagger \Tilde{\textbf{e}}_{0}\rangle),
\end{align}
\begin{align}
    \frac{d\langle \Tilde{\textbf{e}}_{0}^\dagger \Tilde{\textbf{e}}_{0}\rangle}{dt}=
    i\sqrt{N}\xi(\langle a^\dagger \Tilde{\textbf{e}}_{0}\rangle-\langle \Tilde{\textbf{e}}_{0}^\dagger a\rangle) \nonumber\\
    +i\sqrt{N}J(\langle c_p^\dagger \Tilde{\textbf{e}}_{0}\rangle-\langle \Tilde{\textbf{e}}_{0}^\dagger c_p\rangle),
\end{align}
\begin{align}
    \frac{d\langle \Tilde{\textbf{e}}_{0}^\dagger a\rangle}{dt}=-i(\Delta-2g)\langle \Tilde{\textbf{e}}_{0}^\dagger a\rangle+i\sqrt{N}\xi(\langle a^\dagger a\rangle-\langle \Tilde{\textbf{e}}_{0}^\dagger \Tilde{\textbf{e}}_{0}\rangle)\nonumber\\
    +i\sqrt{N}J\langle c_p^\dagger a\rangle,
\end{align}
\begin{align}
    \frac{d\langle \Tilde{\textbf{e}}_{0}^\dagger c_p\rangle}{dt}=-i(\delta-2g)\langle \Tilde{\textbf{e}}_{0}^\dagger c_p\rangle+i\sqrt{N}J(\langle c_p^\dagger c_p\rangle-\langle \Tilde{\textbf{e}}_{0}^\dagger \Tilde{\textbf{e}}_{0}\rangle)\nonumber\\
    +i\sqrt{N}\xi\langle a^\dagger c_p\rangle,
\end{align}
\begin{align}
    \frac{d\langle a^\dagger c_p\rangle}{dt}=i(\Delta-\delta)\langle a^\dagger c_p\rangle+i\sqrt{N}\xi\langle \Tilde{\textbf{e}}_{0}^\dagger c_p\rangle-i\sqrt{N}J\langle a^\dagger\Tilde{\textbf{e}}_{0}\rangle,
\end{align}
\end{subequations}
where $\delta=\omega_p-\omega_d$ and $\Delta=\omega_a-\omega_d$. We have used the following anti-commutation relations
\begin{align}
    \{a,a^\dagger\}=1,\quad\{e_i,e_{j}^\dagger\}=\delta_{ij},\quad\{c_p,c_p^\dagger\}=1.
\end{align}
Also, we note that the $k\neq0$ donor-ring modes do not appear in the set of equations.

\subsection{\label{sec:incoh}Coherent and Incoherent Evolution of Collective Mode Operators}
With use of the Lindblad master equation, we can add dissipation and dephasing to our system. This transforms the equation of motion of $\dot{\rho}$ in Eq.~(\ref{eq:rhodot}) to 
\begin{align}
\label{eq:lbfulll}
    \dot{\rho}=-i[\Tilde{H},\rho]+\mathcal{L}_{D}(\rho)+\mathcal{L}_{A}(\rho)+\mathcal{L}_{P}(\rho),
\end{align}
where $\mathcal{L}_{D}$ describes the dissipation via spontaneous emission from the donors to the environment, $\mathcal{L}_{A}$ describes the charge separation from the acceptor for successful transfer, and $\mathcal{L}_{P}$ is a dephasing term of the ring due to its coupling to the environment. These superoperators are defined as follows:
\begin{subequations}
\begin{align}
    \mathcal{L}_{D}(\rho)=&\sum_{j=1}^N\kappa(2e_j\rho e_j^\dagger-\{e_j^\dagger e_j,\rho\}),
\end{align}
\begin{align}
    \mathcal{L}_{A}(\rho)=&\Gamma(2a\rho a^\dagger-\{a^\dagger a,\rho\}),
\end{align}
\begin{align}
    \mathcal{L}_{P}(\rho)=&\sum_{j=1}^N\gamma(2e_j^\dagger e_j\rho e_j^\dagger e_j-\{e_j^\dagger e_j,\rho\}),
\end{align}
\end{subequations}
where $2\kappa$ is the spontaneous decay rate of the population of each donor atom on the ring, $2\Gamma$ is the spontaneous decay rate of the population of the acceptor for successful transfer, and $2\gamma$ is the dephasing rate of each donor atom.

The equation for an arbitrary operator expectation value $\langle O(t)\rangle$ becomes
\begin{align}
    \frac{d\langle O(t)\rangle}{dt}=-i\Tr[O[H,\rho]]+\Tr[O\mathcal{L}_{D}(\rho)]\nonumber+\Tr[O\mathcal{L}_{A}(\rho)]+\Tr[O\mathcal{L}_{P}(\rho)].
\end{align}
As $c_p$ and $c_p^\dagger$ do not appear in the superoperators, $\mathcal{L}_{D}$, $\mathcal{L}_{A}$ and $\mathcal{L}_{P}$, the equation of motion for $\langle c_p^\dagger c_p\rangle$ is the same as Eq.~(\ref{eq:appcpc}). For $\langle a^\dagger a\rangle$, we find
\begin{align}
    \Tr[a^\dagger a\mathcal{L}_A]=-2\Gamma\langle a^\dagger a\rangle,\quad\Tr[a^\dagger a\mathcal{L}_D]=\Tr[a^\dagger a\mathcal{L}_P]=0,
\end{align}
which gives the full equation of motion as
\begin{align}
    \frac{d\langle a^\dagger a\rangle}{dt}=-2\Gamma\langle a^\dagger a\rangle+i\sqrt{N}\xi(\langle \Tilde{\textbf{e}}_{0}^\dagger a\rangle-\langle a^\dagger \Tilde{\textbf{e}}_{0}\rangle).
\end{align}
For $\langle a^\dagger c_p\rangle$, we find the traces
\begin{align}
    \Tr[a^\dagger c_p\mathcal{L}_A]=-\Gamma\langle a^\dagger c_p\rangle\nonumber,\\
    \Tr[a^\dagger c_p\mathcal{L}_D]=\Tr[a^\dagger c_p\mathcal{L}_P]=0,
\end{align}
which gives the equation of motion for $\langle a^\dagger c_p\rangle$ and, by taking Hermitian conjugate, $\langle c_p^\dagger a\rangle$, as
\begin{align}
    \frac{d\langle a^\dagger c_p\rangle}{dt}=\left(i(\Delta-\delta)-\Gamma\right)\langle a^\dagger c_p\rangle\nonumber\\
    +i\sqrt{N}\xi\langle \Tilde{\textbf{e}}_{0}^\dagger c_p\rangle-i\sqrt{N}J\langle a^\dagger\Tilde{\textbf{e}}_{0}\rangle,
\end{align}
\begin{align}
    \frac{d\langle c_p^\dagger a\rangle}{dt}=\left(i(\delta-\Delta)-\Gamma\right)\langle c_p^\dagger a\rangle\nonumber\\
    +i\sqrt{N}J\langle \Tilde{\textbf{e}}_{0}^\dagger a\rangle-i\sqrt{N}\xi\langle c_p^\dagger\Tilde{\textbf{e}}_{0}\rangle.
\end{align}

To calculate the trace terms for the momentum space operators, we need to first calculate them for the real space operators, as follows
\begin{subequations}
\begin{align}
    &\Tr[a^\dagger e_j\mathcal{L}_A]=-\Gamma\langle a^\dagger e_j\rangle,
\end{align}
\begin{align}
    &\Tr[a^\dagger e_j\mathcal{L}_D]=-\kappa\langle a^\dagger e_j\rangle,
\end{align}
\begin{align}
    &\Tr[a^\dagger e_j\mathcal{L}_P]=-\gamma\langle a^\dagger e_j\rangle,
\end{align}
\end{subequations}
These are then used to calculate trace terms for $a^\dagger\Tilde{\textbf{e}}_k$ as follows
\begin{subequations}
\begin{align}
    \Tr[a^\dagger\Tilde{\textbf{e}}_{0}\mathcal{L}_A]=\frac{1}{\sqrt{N}}\sum_{j=1}^N\Tr[a^\dagger e_j\mathcal{L}_D]\nonumber\\
    =-\Gamma\frac{1}{\sqrt{N}}\sum_{j=1}^N \langle a^\dagger e_j\rangle=-\Gamma\langle a^\dagger\Tilde{\textbf{e}}_{0}\rangle,&
\end{align}
\begin{align}
    \Tr[a^\dagger\Tilde{\textbf{e}}_{0}\mathcal{L}_D]=\frac{1}{\sqrt{N}}\sum_{j=1}^N\Tr[a^\dagger e_j\mathcal{L}_D]\nonumber\\
    =-\kappa\frac{1}{\sqrt{N}}\sum_{j=1}^N \langle a^\dagger e_j\rangle=-\kappa\langle a^\dagger\Tilde{\textbf{e}}_{0}\rangle,&
\end{align}
\begin{align}
    \Tr[a^\dagger\Tilde{\textbf{e}}_{0}\mathcal{L}_D]=\frac{1}{\sqrt{N}}\sum_{j=1}^N\Tr[a^\dagger e_j\mathcal{L}_D]\nonumber\\
    =-\gamma\frac{1}{\sqrt{N}}\sum_{j=1}^N \langle a^\dagger e_j\rangle=-\gamma\langle a^\dagger\Tilde{\textbf{e}}_{0}\rangle.&
\end{align}
\end{subequations}
Which combined with the Hamiltonian term in Eq.~(\ref{eq:nolindblad}), gives
\begin{align}
    \frac{d\langle a^\dagger\Tilde{\textbf{e}}_{0}\rangle}{dt}=(i(\Delta-2g)-(\gamma+\Gamma+\kappa))\langle a^\dagger\Tilde{\textbf{e}}_{0}\rangle\nonumber\\
    +i\sqrt{N}\xi(\langle \Tilde{\textbf{e}}_{0}^\dagger \Tilde{\textbf{e}}_{0}\rangle-\langle a^\dagger a\rangle)-i\sqrt{N}J\langle a^\dagger c_p\rangle,
\end{align}
and by taking the Hermitian conjugate, we obtain the equation for $\langle\Tilde{\textbf{e}}_{0}^\dagger a\rangle$ as
\begin{align}
    \frac{d\langle \Tilde{\textbf{e}}_{0}^\dagger a\rangle}{dt}=(-i(\Delta-2g)-(\gamma+\Gamma+\kappa))\langle \Tilde{\textbf{e}}_{0}^\dagger a\rangle\nonumber\\
    +i\sqrt{N}\xi(\langle a^\dagger a\rangle-\langle \Tilde{\textbf{e}}_{0}^\dagger \Tilde{\textbf{e}}_{0}\rangle)+i\sqrt{N}J\langle c_p^\dagger a\rangle.
\end{align}
Noting that $\Tr[\Tilde{\textbf{e}}_{0}^\dagger c_p\mathcal{L}_A(\rho)]=0$, the same argument can be applied to find the equations for $\langle \Tilde{\textbf{e}}_{0}^\dagger c_p\rangle$ and $\langle c_p^\dagger\Tilde{\textbf{e}}_{0}\rangle$ as
\begin{align}
    \frac{d\langle \Tilde{\textbf{e}}_{0}^\dagger c_p\rangle}{dt}=\left(-i(\delta-2g)-(\gamma+\kappa)\right)\langle \Tilde{\textbf{e}}_{0}^\dagger c_p\rangle\nonumber\\
    +i\sqrt{N}J(\langle c_p^\dagger c_p\rangle-\langle \Tilde{\textbf{e}}_{0}^\dagger \Tilde{\textbf{e}}_{0}\rangle)+i\sqrt{N}\xi\langle a^\dagger c_p\rangle,
\end{align}
\begin{align}
    \frac{d\langle c_p^\dagger\Tilde{\textbf{e}}_{0}\rangle}{dt}=\left(i(\delta-2g)-(\gamma+\kappa)\right)\langle \Tilde{\textbf{e}}_{0}^\dagger c_p\rangle\nonumber\\
    +i\sqrt{N}J(\langle \Tilde{\textbf{e}}_{0}^\dagger \Tilde{\textbf{e}}_{0}\rangle-\langle c_p^\dagger c_p\rangle)-i\sqrt{N}\xi\langle c_p^\dagger a\rangle.
\end{align}

Finally, the trace terms involving $e_j^\dagger e_{j'}$ are
\begin{subequations}
\begin{align}
    &\Tr[e_j^\dagger e_{j'}\mathcal{L}_A]=0,
\end{align}
\begin{align}
    &\Tr[e_j^\dagger e_{j'}\mathcal{L}_D]=-2\kappa\langle e_j^\dagger e_{j'}\rangle,
\end{align}
\begin{align}
\label{eq:dephess}
    &\Tr[e_j^\dagger e_{j'}\mathcal{L}_P]=-2\gamma\langle e_j^\dagger e_{j'}\rangle(1-\delta_{jj'}).
\end{align}
\end{subequations}
Notably, the dephasing term in Eq.~(\ref{eq:dephess}) does not effect the populations, $\langle e_j^\dagger e_{j}\rangle$, but only the coherence terms, $\langle e_j^\dagger e_{j'}\rangle$. If we now consider the $k$-space operators in Eq.~(\ref{eq:kspace}), we can calculate the trace terms
\begin{subequations}
\label{eq:eeee}
\begin{align}
    \Tr[\Tilde{\textbf{e}}_{k}^\dagger\Tilde{\textbf{e}}_{k}\mathcal{L}_D]=\frac{1}{N}\sum_{j,j'}e^{ik(j'-j)}\Tr[e_j^\dagger e_{j'}\mathcal{L}_D]\nonumber\\
    =-\frac{2\kappa}{N}\sum_{j,j'}e^{ik(j'-j)}\langle e_j^\dagger e_{j'}\rangle=-2\kappa\langle\Tilde{\textbf{e}}_{k}^\dagger\Tilde{\textbf{e}}_{k}\rangle,
\end{align}
\begin{align}   
    \Tr[\Tilde{\textbf{e}}_{k}^\dagger\Tilde{\textbf{e}}_{k}\mathcal{L}_P]=\frac{1}{N}\sum_{j,j'}e^{ik(j'-j)}\Tr[e_j^\dagger e_{j'}\mathcal{L}_P]\nonumber\\
    =-\frac{2\gamma}{N}\sum_{j,j'}e^{ik(j'-j)}\langle e_j^\dagger e_{j'}\rangle(1-\delta_{j,j'})\nonumber\\
    =-\frac{2\gamma}{N}\left(\sum_{j,j'}e^{ik(j'-j)}\langle e_j^\dagger e_{j'}\rangle-\sum_j\langle e_j^\dagger e_{j}\rangle\right)\nonumber\\
    =-2\gamma\left(\langle\Tilde{\textbf{e}}_{k}^\dagger\Tilde{\textbf{e}}_{k}\rangle-\frac{1}{N}\sum_{k'}\langle\Tilde{\textbf{e}}_{k'}^\dagger\Tilde{\textbf{e}}_{k'}\rangle\right),
\end{align}
\end{subequations}
where we have used
\begin{align}
    \sum_{j=1}^N\langle e_j^\dagger e_{j}\rangle=\sum_{k}\langle\Tilde{\textbf{e}}_{k}^\dagger\Tilde{\textbf{e}}_{k}\rangle.
\end{align}
For $k=0$, these equations become
\begin{subequations}
\begin{align}
    \Tr[\Tilde{\textbf{e}}_{0}^\dagger\Tilde{\textbf{e}}_{0}\mathcal{L}_D]=-2\kappa\langle\Tilde{\textbf{e}}_{0}^\dagger\Tilde{\textbf{e}}_{0}\rangle,&
\end{align}
\begin{align}   
    \Tr[\Tilde{\textbf{e}}_{0}^\dagger\Tilde{\textbf{e}}_{0}\mathcal{L}_P]=-2\gamma\left((1-\frac{1}{N})\langle\Tilde{\textbf{e}}_{0}^\dagger\Tilde{\textbf{e}}_{0}\rangle-\frac{1}{N}\langle\Tilde{\textbf{e}}_{k}^\dagger\Tilde{\textbf{e}}_{k}\rangle_L\right),&
\end{align}
\end{subequations}
where we have introduced $\langle\Tilde{\textbf{e}}_{k}^\dagger\Tilde{\textbf{e}}_{k}\rangle_L$, the sum of the expectations of the $k\neq0$ populations operators,
\begin{eqnarray}
    \langle\Tilde{\textbf{e}}_{k}^\dagger\Tilde{\textbf{e}}_{k}\rangle_L=\sum_{k\neq0}\langle\Tilde{\textbf{e}}_{k}^\dagger\Tilde{\textbf{e}}_{k}\rangle.
\end{eqnarray}
This gives the equation of motion for $\langle\Tilde{\textbf{e}}_{0}^\dagger\Tilde{\textbf{e}}_{0}\rangle$, as
\begin{align}
\label{eq:b18}
    \frac{d\langle \Tilde{\textbf{e}}_{0}^\dagger \Tilde{\textbf{e}}_{0}\rangle}{dt}=-2\left(\kappa+\gamma\left(1-\frac{1}{N}\right)\right)\langle \Tilde{\textbf{e}}_{0}^\dagger \Tilde{\textbf{e}}_{0}\rangle+\frac{2\gamma}{N}\langle\Tilde{\textbf{e}}_{k}^\dagger\Tilde{\textbf{e}}_{k}\rangle_L\nonumber\\
    +i\sqrt{N}\xi(\langle a^\dagger \Tilde{\textbf{e}}_{0}\rangle-\langle \Tilde{\textbf{e}}_{0}^\dagger a\rangle) +i\sqrt{N}J(\langle c_p^\dagger \Tilde{\textbf{e}}_{0}\rangle-\langle \Tilde{\textbf{e}}_{0}^\dagger c_p\rangle),
\end{align}
In Eq.~(\ref{eq:b18}) we need to determine the equation of motion for $\langle \Tilde{\textbf{e}}_{k}^\dagger \Tilde{\textbf{e}}_{k}\rangle_L$, since we already have the equations of motion for the other operators in Eq.~(\ref{eq:nolindblad}). Since only the $k=0$ mode couples to the acceptor, we observe that for $k\neq0$,
\begin{align}
    \Tr[-i\Tilde{\textbf{e}}_{k}^\dagger\Tilde{\textbf{e}}_{k}[H,\rho]]=0.
\end{align}
From Eq.~(\ref{eq:eeee}) we obtain the following
\begin{subequations}
\begin{align}
    \sum_{k\neq0}\Tr[\Tilde{\textbf{e}}_{k}^\dagger\Tilde{\textbf{e}}_{k}\mathcal{L}_D]=-2\kappa\sum_{k\neq0}\langle\Tilde{\textbf{e}}_{k}^\dagger\Tilde{\textbf{e}}_{k}\rangle=-2\kappa\langle\Tilde{\textbf{e}}_{k}^\dagger\Tilde{\textbf{e}}_{k}\rangle_L,&
\end{align}
\begin{align}   
    \sum_{k\neq0}\Tr[\Tilde{\textbf{e}}_{k}^\dagger\Tilde{\textbf{e}}_{k}\mathcal{L}_P]=&2\gamma\left(\frac{N-1}{N}\langle\Tilde{\textbf{e}}_{0}^\dagger\Tilde{\textbf{e}}_{0}\rangle-\frac{1}{N}\sum_{k\neq0}\langle\Tilde{\textbf{e}}_{k}^\dagger\Tilde{\textbf{e}}_{k}\rangle\right)\nonumber\\
    =&2\gamma\left(\frac{N-1}{N}\langle\Tilde{\textbf{e}}_{0}^\dagger\Tilde{\textbf{e}}_{0}\rangle-\frac{1}{N}\langle\Tilde{\textbf{e}}_{k}^\dagger\Tilde{\textbf{e}}_{k}\rangle_L\right),
\end{align}
\end{subequations}
which combine to give
\begin{align}
    \frac{d\langle\Tilde{\textbf{e}}_{k}^\dagger\Tilde{\textbf{e}}_{k}\rangle_L}{dt}=2\left(-\kappa-\frac{\gamma}{N}\right)\langle\Tilde{\textbf{e}}_{k}^\dagger\Tilde{\textbf{e}}_{k}\rangle_L+2\gamma\left(1-\frac{1}{N}\right)\langle \Tilde{\textbf{e}}_{0}^\dagger \Tilde{\textbf{e}}_{0}\rangle.
\end{align}
In matrix form, these equations can be written as
\begin{align}
\label{eq:matform}
    \frac{d\textbf{P}(t)}{dt}=\textbf{M}\textbf{P}(t),
\end{align}
where
\begin{widetext}
\begin{subequations}
\begin{align}
\label{eq:appendpvec}
    \textbf{P}(t)=\left(
\begin{smallmatrix}\langle c_p^\dagger c_p(t)\rangle&\langle a^\dagger a(t)\rangle&\langle \Tilde{\textbf{e}}_{0}^\dagger \Tilde{\textbf{e}}_{0}(t)\rangle&\langle \Tilde{\textbf{e}}_{0}^\dagger c_p(t)\rangle&\langle c_p^\dagger\Tilde{\textbf{e}}_{0}(t)\rangle&\langle a^\dagger c_p(t)\rangle&\langle c_p^\dagger a(t)\rangle&\langle \Tilde{\textbf{e}}_{0}^\dagger a(t)\rangle&\langle a^\dagger\Tilde{\textbf{e}}_{0}(t)\rangle&\langle\Tilde{\textbf{e}}_{k}^\dagger\Tilde{\textbf{e}}_{k}(t)\rangle_L\end{smallmatrix}\right)^T,
\end{align}
\begin{align}
\label{eq:appenmat}
    \textbf{M}=\left(
\begin{smallmatrix}0 & 0 & 0 & i J \sqrt{N} & -i J \sqrt{N} & 0 & 0 & 0 & 0 & 0 \\
 0 & -2 \Gamma & 0 & 0 & 0 & 0 & 0 & -i \sqrt{N} \xi  & i \sqrt{N} \xi  & 0 \\
 0 & 0 & -2\left(\kappa+\gamma\left(1-\frac{1}{N}\right)\right) & -i J \sqrt{N} & i J \sqrt{N} & 0 & 0 & i \sqrt{N} \xi  & -i
   \sqrt{N} \xi  & \frac{2 \gamma }{N} \\
 i J \sqrt{N} & 0 & -i J \sqrt{N} & -i \delta_c -\gamma -\kappa  & 0 & i \sqrt{N} \xi  & 0 & 0 & 0 & 0 \\
 -i J \sqrt{N} & 0 & i J \sqrt{N} & 0 & i \delta_c -\gamma -\kappa  & 0 & -i \sqrt{N} \xi  & 0 & 0 & 0 \\
 0 & 0 & 0 & i \sqrt{N} \xi  & 0 & i (\Delta-\delta )-\Gamma & 0 & -i J \sqrt{N} & 0 & 0 \\
 0 & 0 & 0 & 0 & -i \sqrt{N} \xi  & 0 & i (\delta -\Delta)-\Gamma & 0 & i J \sqrt{N} & 0 \\
 0 & -i \sqrt{N} \xi  & i \sqrt{N} \xi  & 0 & 0 & -i J \sqrt{N} & 0 & i \Delta_c-\gamma -\kappa -\Gamma &
   0 & 0 \\
 0 & i \sqrt{N} \xi  & -i \sqrt{N} \xi  & 0 & 0 & 0 & i J \sqrt{N} & 0 & -i \Delta_c-\gamma -\kappa
   -\Gamma & 0 \\
 0 & 0 & \frac{2 \gamma  (N-1)}{N} & 0 & 0 & 0 & 0 & 0 & 0 & -\frac{2\gamma}{N}-2\kappa  \\
\end{smallmatrix}\right),
\end{align}
\end{subequations}
where $\delta_c=\delta-2g$ and $\Delta_c=\Delta-2g$.
\end{widetext}

\subsection{\label{app:transf}A Derivation of Transfer Efficiency Formula}
The transfer efficiency, $\eta$, is the probability for a successful excitation transfer and is defined as
\begin{align}
\label{eq:etadeff}
    \eta=\int_0^{\infty}2\Gamma\langle a^\dagger a(t)\rangle dt.
\end{align}
To obtain this, we could solve Eq.~(\ref{eq:matform}) to find $\langle a^\dagger a(t)\rangle$ and then integrate to obtain $\eta$. Here we present an alternate method, whereby we use Laplace transformations to obtain $\eta$ without directly solving for $\langle a^\dagger a(t)\rangle$.

The Laplace transformation $F(s)$ of an arbitrary function $f(t)$, is defined as
\begin{align}
    F(s)=\textbf{L}[f(t)]_s=\int_0^{\infty} e^{-st}f(t)dt,
\end{align}
where $s$ is a complex parameter. The Laplace transform of the time derivative of a function is
\begin{align}
    \textbf{L}\left[\frac{df(t)}{dt}\right]_s=sF(s)-f(0).
\end{align}
For a system of equations that can be written in the following form
\begin{align}
\label{eq:c3ap}
    \frac{d\textbf{P}(t)}{dt}=\textbf{M}\textbf{P}(t),
\end{align}
where $\textbf{P}(t)$ is a column vector containing each operator expectation value, and $\textbf{M}$ is a square matrix of dimension $m$, the Laplace transform of Eq.~(\ref{eq:c3ap}) is given as
\begin{align}
    s\Tilde{\textbf{P}}(s)-\textbf{P}(0)=\textbf{M}\Tilde{\textbf{P}}(s),
\end{align}
where $\Tilde{\textbf{P}}(s)$ is the Laplace transform of $\textbf{P}(t)$. This can be rearranged as 
\begin{align}
\label{eq:c5}
    \textbf{A}(s)\Tilde{\textbf{P}}(s)=-\textbf{P}(0),
\end{align}
where $\textbf{A}(s)=\textbf{M}-s\textbf{I}$ and $\textbf{I}$ is the identity matrix. Pre-multiplying both sides of Eq.~(\ref{eq:c5}) by $\textbf{A}^{-1}(s)$, we obtain
\begin{align}
\label{eq:lapspace}
    \Tilde{\textbf{P}}(s)=-\frac{1}{\det[\textbf{A}(s)]}\textbf{C}^T\textbf{P}(0),
\end{align}
where 
\begin{equation}
    \textbf{A}^{-1}(s)=\frac{\textbf{C}^T}{\det[\textbf{A}(s)]},
\end{equation}
and $\textbf{C}^T$ is the transpose of the co-factor matrix of $\textbf{A}(s)$. The determinant of $\textbf{A}(s)$ is the determinant of $(\textbf{M}-s\textbf{I})$, which is given as
\begin{align}
\label{eq:asdet}
    \det[\textbf{A}(s)]=\prod^{m}_{j=1}(\epsilon^{(j)}_M-s),
\end{align}
where $\epsilon^{(j)}_M$ ($j=\{1,...,m\})$ are the eigenvalues of matrix $\textbf{M}$. 

The initial condition is $\langle c_p^\dagger c_p(0)\rangle=1$. For the ring-acceptor system with a single photon, we consider the vector $\textbf{P}(t)$ and matrix $\textbf{M}$, as defined in Eq.~(\ref{eq:appendpvec}) and (\ref{eq:appenmat}), respectively. Note that the first and second elements of $\textbf{P}(t)$, i.e. $p_1(t)$ and $p_2(t)$, are $\langle c_p^\dagger c_p(t)\rangle$ and $\langle a^\dagger a(t)\rangle$, respectively. $\langle a^\dagger a(t)\rangle$ can be found via an inverse transformation on the second element of $\Tilde{\textbf{P}}(s)$ in Eq.~(\ref{eq:lapspace}), which is
\begin{align}
\label{eq:lapacc}
    \langle a^\dagger a(s)\rangle=-\frac{1}{\det[\textbf{A}(s)]}c_{12}^T.
\end{align}
where $c^{T}_{12}$ is the $(1,2)^{th}$ element of $\textbf{C}^{T}$, which is calculated as
\begin{equation}
    c^{T}_{12}=-\det[\textbf{A}^{M}_{2,1}(s)],
\end{equation}
where $\textbf{A}^{M}_{2,1}(s)$ is the $(2,1)^{th}$ sub-matrix of $\textbf{A}(s)$. It should be noted that the sub-matrix, $\textbf{A}^{M}_{i,j}$, is constructed by eliminating the $i^{th}$ row and $j^{th}$ column from matrix $\textbf{A}$. The determinant of $\textbf{A}^{M}_{2,1}(s)$ gives the following characteristic polynomial
\begin{align}
\label{eq:ampoly}
    \det[\textbf{A}^{M}_{2,1}]=\sum_{j=0}^{9}a_js^j,
\end{align}
where $a_j$ is the coefficient of $s^j$. We note that $\textbf{A}^{M}_{2,1}(s)$ is an $9\times9$ matrix so its determinant is a polynomial of degree $9$. Substituting Eq.~(\ref{eq:asdet}) and~(\ref{eq:ampoly}) into Eq.~(\ref{eq:lapacc}) gives
\begin{align}
\label{eq:aas}
    \langle a^\dagger a(s)\rangle=\frac{\sum_{j=0}^{9}a_js^j}{\prod^{10}_{i=1}(\epsilon^{(i)}_M-s)}.
\end{align}
Applying an inverse Laplace transformation to Eq.~(\ref{eq:aas}) gives $\langle a^\dagger a(t)\rangle$ as (see Appendix~\ref{app:addproof})
\begin{align}
\label{eq:c12}
    \langle a^\dagger a(t)\rangle=\sum_{k=1}^{10}\left(\frac{e^{\epsilon^{(k)}_M t}}{\prod_{i\neq k}(\epsilon^{(k)}_M-\epsilon^{(i)}_M)}\sum_{j=0}^{9}a_j(\epsilon^{(k)}_M)^j\right).
\end{align}
We can substitute the r.h.s. of Eq.~(\ref{eq:c12}) into the integral in Eq.~(\ref{eq:etadeff}) and thereby obtain the transfer efficiency $\eta$. Since the polynomial in Eq.~(\ref{eq:ampoly}) has order less than $N=10$, the $j\neq0$ terms cancel, and we find the efficiency, $\eta$, as 
\begin{align}
    \eta=2\Gamma\frac{a_0}{\prod_{j=1}^{10}\epsilon^{(j)}_M}.
\end{align}
We note that the denominator is the determinant of matrix $\textbf{M}$, and the numerator is the determinant of the $(2,1)^{th}$ sub-matrix of $\textbf{M}$, i.e. $\textbf{M}_{(2,1)}^M$. The efficiency can then be written in the form
\begin{align}
\label{eq:etappen}
    \eta=2\Gamma\frac{\text{Det}[\textbf{M}_{(2,1)}^M]}{\text{Det}[\textbf{M}]}.
\end{align}
We note that the integral only converges if the eigenvalues $\epsilon^{(j)}_M$ all have negative real parts, which is true for a system with dissipation and without driving.

\section{\label{app:addproof}Inverse Laplace Transform of $\langle a^\dagger a(s)\rangle$} 
The Laplace transform $\langle a^\dagger a(s)\rangle$ has form
\begin{align}
\label{eq:d1}
    \langle a^\dagger a(s)\rangle=\frac{Q(s)}{D(s)},
\end{align}
where
\begin{align}
    Q(s)=\sum_{j=0}^{N-1}a_js^j,\quad D(s)=\prod^{N}_{i=1}(\epsilon^{(i)}_M-s).
\end{align}
If we assume there are no degeneracies, $\epsilon^{(i)}_M$ ($i=1,...,N$) are completely distinct and we can rewrite Eq.~(\ref{eq:d1}) in the form
\begin{align}
\label{eq:sumcj}
    \langle a^\dagger a(s)\rangle=\sum_{j=1}^N\frac{C_{j}}{(\epsilon^{(i)}_M-s)}.
\end{align}
By multiplying both sides of Eq.~(\ref{eq:sumcj}) by $(\epsilon^{(j)}_M-s)$ and substituting $s=\epsilon^{(j)}_M$, we find $C_j$ as,
\begin{align}
    C_j=\left[(\epsilon^{(j)}_M-s)\frac{Q(s)}{D(s)}\right]_{s=\epsilon^{(j)}_M},
\end{align}
and we can apply an inverse Laplace transform to Eq.~(\ref{eq:sumcj}) to obtain
\begin{align}
    \langle a^\dagger a(t)\rangle=-\sum_{j=1}^N C_je^{\epsilon^{(i)}_M t},
\end{align}
We can integrate this to find the efficiency
\begin{align}
    \eta=2\Gamma\sum_{j=1}^N \frac{C_j}{\epsilon^{(i)}_M}.
\end{align}
We can simplify this by equating equations Eq.~(\ref{eq:d1}) and~(\ref{eq:sumcj})
to find
\begin{align}
    \frac{\sum_{j=0}^{N-1}a_js^j}{\prod^{N}_{i=1}(\epsilon^{(i)}_M-s)}=\sum_{j=1}^N\frac{C_{j}}{(\epsilon^{(i)}_M-s)},
\end{align}
which holds for all $s$. So if we take $s=0$ we obtain
\begin{align}
\label{eq:nodegen}
    \eta=2\Gamma\sum_{j=1}^N\frac{C_{j}}{\epsilon^{(i)}_M}=2\Gamma\frac{a_0}{\prod^{N}_{i=1}\epsilon^{(i)}_M}.
\end{align}

If there are degeneracies in $\epsilon^{(i)}_M$ ($i=1,...,N$), for example $r$-fold degeneracy of $\epsilon^{(d)}_M$ ($d=N-r+1$), Eq.~(\ref{eq:sumcj}) becomes
\begin{align}
\label{eq:sumkj}
    &\langle a^\dagger a(s)\rangle=\sum_{\substack{j=1\\j\neq d}}^{N-r}\frac{C_{j}}{(\epsilon^{(i)}_M-s)}+\sum_{i=1}^r\frac{k_i}{(\epsilon^{(d)}_M-s)^i},\\
    &C_j=\left[(\epsilon^{(j)}_M-s)\frac{Q(s)}{D(s)}\right]_{s=\epsilon^{(j)}_M},\\
    &k_i=\frac{1}{(r-i)!}\frac{d^{r-i}}{ds^{r-i}}\left[(\epsilon^{(d)}_M-s)^r\frac{Q(s)}{D(s)}\right]_{s=\epsilon^{(d)}_M},
\end{align}
where the first term in Eq.~(\ref{eq:sumkj}) gives the contribution from the non-degenerate eigenvalues, and the second gives the contribution from the degenerate eigenvalues. An inverse Laplace transformation on Eq.~(\ref{eq:sumkj}) gives
\begin{align}
\label{eq:d12}
    \langle a^\dagger a(t)\rangle=-\sum_{\substack{j=1\\j\neq d}}^{N-r}C_je^{\epsilon^{(i)}_M t}+\sum_{i=1}^r\frac{(-1)^ik_i t^{(i-1)}}{(i-1)!}e^{\epsilon^{(d)}_M t},
\end{align}
which we integrate to obtain the efficiency as
\begin{align}
    \eta=2\Gamma\left(\sum_{\substack{j=1\\j\neq d}}^{N-r}\frac{C_j}{\epsilon^{(i)}_M}+\sum_{i=1}^r\frac{k_i}{(\epsilon^{(d)}_M)^i}\right).
\end{align}
In integrating Eq.~(\ref{eq:d12}) we have used the following
\begin{align}
    \int_{0}^{\infty}\frac{t^{n}}{n!}e^{a t}dt=\left(\frac{-1}{a}\right)^n,
\end{align}
for $n=0,1,2,...$ and $\Re[a]<0$. Now we can use the same technique as before, by equating Eq.~(\ref{eq:d1}) and Eq.~(\ref{eq:sumkj}), for $s=0$, we find
\begin{align}
\label{eq:degenn}
    \eta=2\Gamma\sum_{\substack{j=1\\j\neq d}}^{N-r}\frac{C_j}{\epsilon^{(i)}_M}+\sum_{i=1}^r\frac{k_i}{(\epsilon^{(d)}_M)^i}=2\Gamma\frac{a_0}{\prod^{N}_{i=1}\epsilon^{(i)}_M},
\end{align}
where $a_0$ is defined in Eq.~(\ref{eq:ampoly}). It should be noted that the result in Eq.~(\ref{eq:degenn}) with degeneracies is the same as the result in Eq.~(\ref{eq:nodegen}) without degeneracies.

\bibliography{apssamp}
\end{document}